\documentclass[aps,floatfix,longbibliography,footinbib,notitlepage,showpacs]{revtex4-2}
\usepackage{physics}
\usepackage{graphicx}
\usepackage[T1]{fontenc}
\usepackage{amsthm,amsmath,amsfonts,dsfont,color,xcolor,mathrsfs,mathtools,bbm,amssymb}
\usepackage{tikz}
\usetikzlibrary{calc}
\definecolor{figaccent}{HTML}{C04F17}
\definecolor{Bittersweet}{HTML}{C04F17}
\usepackage[bookmarksnumbered,colorlinks,plainpages,linktocpage]{hyperref}
\definecolor{Darkgreen}{rgb}{0,0.4,0}
\hypersetup{%
    pdfborder = {0 0 0},
    colorlinks,
    citecolor=blue,
    filecolor=Darkgreen,
    linkcolor=Bittersweet,
    urlcolor=blue}

\newtheorem{theorem}{Theorem}
\newtheorem{remark}[theorem]{Remark}
\newtheorem{lemma}[theorem]{Lemma}
\newtheorem{proposition}[theorem]{Proposition}
\newtheorem{corollary}[theorem]{Corollary}

\theoremstyle{definition}
\newtheorem{example}[theorem]{Example}
\theoremstyle{plain}

\newcommand{\medotimes}{\mathbin{\raisebox{0.13ex}{\scalebox{0.8}{$\bigotimes$}}}}

\def\ChR{0.34}   
\def\ChW{0.20}   
\def\ChH{0.18}   
\def\ChD{0.88}   
\def\ChS{70}     

\pgfmathsetmacro\ChRw{(0.25+\ChW*\ChW-\ChR*\ChR)/(2*(\ChR-\ChW))}
\pgfmathsetmacro\ChA{atan2(\ChW+\ChRw,0.5)}
\pgfmathsetmacro\ChDr{\ChD*\ChR}

\definecolor{chainred}{RGB}{235,0,0}
\tikzset{
  outer plate/.style={fill=black,  draw=chainred, line width=1.3pt, line join=round},
  inner plate/.style={fill=black!40,draw=chainred, line width=1.3pt, line join=round},
  hidden edge/.style={draw=black!22, line width=0.45pt, dash pattern=on 1.6pt off 1.1pt},
  pin/.style={fill=black!3},
}

\def\ChPlate{%
  (\ChA:\ChR)
  arc[start angle=180+\ChA, end angle=360-\ChA, radius=\ChRw]
  arc[start angle=180-\ChA, end angle=\ChA-180,  radius=\ChR]
  arc[start angle=\ChA,     end angle=180-\ChA,  radius=\ChRw]
  arc[start angle=-\ChA,    end angle=\ChA-360,  radius=\ChR] -- cycle}

\def\ChTailLink{0}\def\ChTailA{0}\def\ChTailB{0}
\newcommand\chain[3]{%
  \xdef\ChX{#1}\xdef\ChY{#2}\gdef\ChList{}%
  \foreach \h [count=\i] in {#3}{%
    \xdef\ChList{\ChList\ifnum\i>1 ,\fi\ChX/\ChY/\h/\i}%
    \pgfmathsetmacro\nx{\ChX+cos(\h)}\pgfmathsetmacro\ny{\ChY+sin(\h)}%
    \xdef\ChX{\nx}\xdef\ChY{\ny}}%
  \foreach \X/\Y/\h/\i in \ChList{\ifodd\i\else
    \path[inner plate, shift={(\X,\Y)}, rotate=\h] \ChPlate;\fi}%
  \foreach \X/\Y/\h/\i in \ChList{\ifodd\i
    \path[outer plate, shift={(\X,\Y)}, rotate=\h] \ChPlate;\fi}%
  \foreach \X/\Y/\h/\i in \ChList{\ifodd\i\else
    \begin{scope}[shift={(\X,\Y)}, rotate=\h]
      \path[hidden edge] (180-\ChS:\ChDr)
        arc[start angle=180-\ChS, end angle=180+\ChS, radius=\ChDr];
      \ifnum\i=\ChTailLink
        \path[hidden edge] ($(1,0)+(\ChTailA:\ChDr)$)
          arc[start angle=\ChTailA, end angle=\ChTailB, radius=\ChDr];
      \else
        \path[hidden edge] ($(1,0)+(-\ChS:\ChDr)$)
          arc[start angle=-\ChS, end angle=\ChS, radius=\ChDr];
      \fi
    \end{scope}\fi}%
  \foreach \X/\Y/\h/\i in \ChList{\path[pin] (\X,\Y) circle[radius=\ChH];}%
  \path[pin] (\ChX,\ChY) circle[radius=\ChH];
}
\begin{document}

\hyphenpenalty=10000
\exhyphenpenalty=10000
\sloppy
\title{Entanglement Geometry Beyond Polygons:\\Graph-Theoretic Entanglement Inequalities}

\author{Masoud Gharahi}
\email[]{masoud.gharahi@gmail.com}
\email[]{masoud.gharahighahi@uj.edu.pl}
\affiliation{Faculty of Physics, Astronomy and Applied Computer Science, Jagiellonian University, 30-348 Kraków, Poland}
\begin{abstract}
Entanglement entropies of pure multipartite states can be constrained by organizing subsystem cuts as edges (hyperedges) of graphs (hypergraphs). Starting from the one-party entanglement polygon inequality, we derive these constraints from strong subadditivity and pure-state complementarity, with optimal coefficients and explicit equality conditions. Our central result determines, for a distinguished two-party cut and an arbitrary weighted family of cuts, the optimal polygon coefficient: the smallest total weight of the cuts separating a member of the distinguished pair from a remaining party. A refinement involving three mutual information terms gives the exact minimum of the weighted sum at fixed entropies of the two parties and of their union, attained by at most three independent bipartite pure states. The analogous coefficient is also optimal for a distinguished one-party cut and for any distinguished cut that the family does not split. This criterion recovers the cycle and complete-graph inequalities and shows that uniform incidence parity, which we prove sufficient for polygon inequalities on hypergraphs, is not necessary. Averaging strong subadditivity over both sides of a bipartition gives a sharp $k$-marginal inequality with coefficient $2\binom{n-2}{k-1}$, and we determine uniformly optimal coefficients for collective hyperstar inequalities. For paths and overlapping hyperchains, the deficits are exact sums of conditional mutual information terms; genuinely multipartite entangled cluster states saturate the hyperchain inequality while the chain inequality remains strict. We characterize pure chain equality states, bound the trace distance to cycle equality states independently of the number of parties and local dimensions, and obtain bond-dimension bounds for matrix product states and sequential-recovery error estimates.
\end{abstract}
\maketitle
\section{Introduction}\label{sec:introduction}

\begin{figure}[ht]
\centering
  \resizebox{0.6\linewidth}{!}{%
  \begin{tikzpicture}
    {\def\ChTailLink{14}\def\ChTailA{-110}\def\ChTailB{-25}%
    \chain{0}{0}{90,112.5,135,157.5,180,202.5,225,247.5,270,292.5,315,337.5,360,382.5,315}}
    \chain{6.93}{0}{90,112.5,135,157.5,180,202.5,225,247.5,270,292.5,315,337.5,360,382.5,405,427.5}
    \begin{scope}[shift={(-5.28,-3.95)}, scale=0.83]
      \chain{0}{0}{0,0,0,0,0,0,0,0,0,0,0,0,0,0,0}
    \end{scope}
  \end{tikzpicture}%
  }
\caption{Graphical organization of subsystem cuts underlying multipartite entanglement inequalities. White disks represent the parties $A_j$ of a pure $n$-partite state ($n=16$ shown). Each black or gray link (edge) selects an adjacent pair $A_jA_{j+1}$, with $\mathcal{E}_{A_jA_{j+1}}$ denoting its entanglement with the complementary subsystem. Removing one edge from the cycle (upper right) yields an open chain (upper left), shown in straightened form below. The path and cycle geometries underlie the Chain Entanglement Inequality (CEI) and the Edge Entanglement Inequality (EEI), respectively.}
\label{fig:path-cycle}
\end{figure}
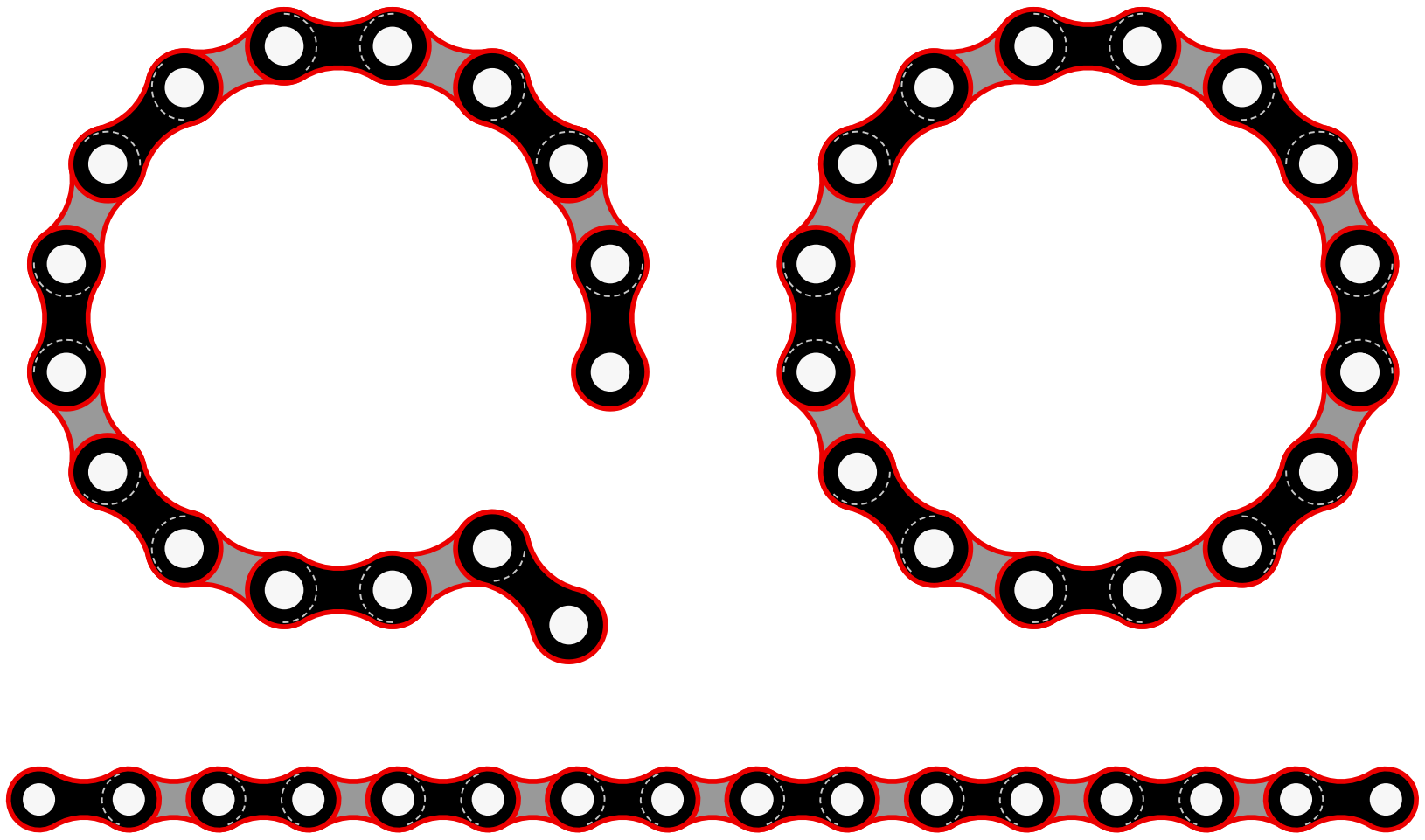

Quantum entanglement obeys fundamental restrictions on how it can be shared among different subsystems. Understanding these restrictions is a central problem of quantum information theory. Such constraints are relevant to quantum communication \cite{BBCJPW93}, cryptography \cite{Ekert91}, computation \cite{NC00}, and many-body physics \cite{AFOV08}, where the distribution of entanglement influences the properties and uses of multipartite states.

A familiar restriction is entanglement monogamy. For a pure three-qubit state, the Coffman--Kundu--Wootters (CKW) inequality reads \cite{CKW00}
\begin{equation}\label{intro:three-party-CKW}
\mathcal{C}_{A_1\mid A_2A_3}^{2} \geq
\mathcal{C}_{A_1\mid A_2}^{2}
+\mathcal{C}_{A_1\mid A_3}^{2},
\end{equation}
where $\mathcal{C}$ denotes concurrence \cite{Wootters98}. The term on the left concerns the pure bipartition $A_1\,|\,A_2A_3$; the two terms on the right are evaluated on the mixed reductions $\varrho_{A_1A_2}$ and $\varrho_{A_1A_3}$. This relation extends to arbitrary multiqubit systems \cite{OV06}, while the same form can fail in higher local dimensions \cite{Ou07}.

The von Neumann entropy provides a way to study entanglement across pure-state bipartitions \cite{Neumann}. For a pure global state, the entanglement entropy across a bipartition equals the entropy of either side. Standard entropy inequalities---subadditivity (SA), the Araki--Lieb inequality, weak monotonicity (WM), and strong subadditivity (SSA) \cite{AL70,LR73}---therefore constrain the entanglement entropies of different subsystem cuts. Our derivations use SA, WM, and SSA, together with
pure-state complementarity. The graph and hypergraph
constructions specify the subsystem cuts to which these
properties are applied.

The entanglement polygon inequality (EPI) provides a starting point. Polygon constraints were established for the spectra of one-qubit reduced states \cite{HSS03}; entanglement polygon relations were subsequently studied for qubit and qudit systems \cite{QAE18,YYL22,Shi23}. For a pure state $|\psi\rangle$ on $XY$, write $\mathsf{E}_{X\mid Y}(|\psi\rangle) =\mathcal{S}(\varrho_X)=\mathcal{S}(\varrho_Y)$ for its entanglement entropy across $X\,|\,Y$. For a pure state of three parties $A_1,A_2,A_3$, the entropy-based EPI takes the form
\begin{equation}\label{intro:three-party-EPI}
\mathsf{E}_{A_1\mid A_2A_3}\leq
\mathsf{E}_{A_2\mid A_1A_3}
+\mathsf{E}_{A_3\mid A_1A_2},
\end{equation}
together with its permutations. For more parties, each one-party entanglement is bounded by the sum of the others. These entropy bounds follow from SA and the equality of complementary entropies for a pure state. CKW relates concurrence across a pure-state cut to pairwise concurrence within mixed reductions, whereas the entropy-based EPI compares bipartitions of the same global pure state.

In what follows, $I_k$ selects $k$ parties and $\bar{I}_k$ denotes the remaining parties. For a pure global state $|\psi\rangle$, we write $\mathcal{E}_{I_k}\coloneqq\mathsf{E}_{I_k\mid\bar{I}_k}(|\psi\rangle)=\mathcal{S}(\varrho_{I_k})$. Thus $\mathcal{E}_{A_iA_j}$ treats $A_iA_j$ as one composite subsystem.

We begin the extension beyond one-party cuts with the Marginal Entanglement Inequality (MEI), which bounds the entanglement of any subsystem by one half of the total one-party entanglement. The Strong Marginal Entanglement Inequality (SMEI) gives separate bounds from the one-party entropies on the two sides of the bipartition. These comparisons allow the selected subsystems to contain more than one party.

The arrangement of the selected cuts also matters. On a path, the Chain Entanglement Inequality (CEI) compares adjacent two-party cuts with the interior one-party cuts. Its deficit is exactly a sum of conditional mutual information terms, so equality holds precisely when every term vanishes. In entropic form, this inequality is an instance of the Markov entropy decomposition~\cite{PH11}; here it is read as a constraint between entanglement cuts, with new consequences. Closing the path gives the cycle-edge geometry and the Edge Entanglement Inequality (EEI), which compares a distinguished cycle edge with the other cycle edges. The two arrangements are illustrated in Fig.~\ref{fig:path-cycle}.

Including every pair of parties as an edge leads to the Two-Marginal Entanglement Polygon Inequality (TMEPI). Let $\mathfrak{I}_k$ denote all distinct choices of $k$ parties, each counted once, and let $\mathcal{E}_{\mathfrak{I}_k}\coloneqq\sum_{I_k\in\mathfrak{I}_k}\mathcal{E}_{I_k}$. Writing $\overline{\mathcal{E}}_k\coloneqq\binom{n}{k}^{\!-1}\mathcal{E}_{\mathfrak{I}_k}$, the average marginal entanglement hierarchy reads
\begin{equation}\label{intro:average-hierarchy}
\overline{\mathcal{E}}_1
\leq\overline{\mathcal{E}}_2
\leq\cdots\leq
\overline{\mathcal{E}}_{\lfloor n/2\rfloor}.
\end{equation}
The hierarchy also follows from the symmetrized entropy cone~\cite{Pip03,FHC22}, although, to the best of our knowledge, it is not stated explicitly there; we therefore include a direct proof. An average $k$-marginal entanglement inequality also bounds these means in terms of the one-marginal total. Combining the hierarchy with the MEI gives the $k$-Marginal Entanglement Inequality ($k$-MEI). Averaging SSA separately over the two sides of a distinguished bipartition yields the stronger sharp $k$-MEI,
\begin{equation}\label{intro:sharp-k-MEI}
2\binom{n-2}{k-1}\mathcal{E}_{I_k}
\leq\mathcal{E}_{\mathfrak{I}_k}.
\end{equation}
When expressed in terms of sums, the same numerical coefficient appears in a general set-function inequality of Filmus \cite{Filmus13}, proved there by a different technique. The Generalized Entanglement Polygon Inequality (GEPI) gives the polygon form at arbitrary marginal order. We also establish powered marginal and cyclic inequalities for $\alpha\in(0,1]$.

The passage from graphs to hypergraphs replaces edges by larger subsystem sets and leads to hypergraph inequalities. Consecutive overlapping hyperedges give the Hyperchain Entanglement Inequality (HCEI), whose deficit, like that of the CEI, is an exact sum of conditional mutual information terms. For a family $I^{(1)},\ldots,I^{(q)}$ of nonempty proper subsystems, possibly of different sizes, the Symmetric-Difference Entropy Triangle Inequality provides the key estimate. If all parties have the same incidence parity (all even or all odd), the Parity-Hypergraph Entanglement Polygon Inequality (PHEPI) gives
\begin{equation}\label{intro:parity-hypergraph}
2\,\mathcal{E}_{I^{(s)}}\leq \sum_{t=1}^{q}\mathcal{E}_{I^{(t)}}, \qquad s=1,\ldots,q.
\end{equation}
Its graph specialization is the Parity-Graph Entanglement
Polygon Inequality (PGEPI). Applying the same condition to cyclic
windows gives the Hyperedge Entanglement Inequality (HEEI),
extending the EEI to hypercycles.

The parity condition is sufficient but not necessary. For a distinguished two-party cut, we determine the optimal coefficient exactly for an arbitrary family $\mathcal{F}$ of cuts with nonnegative weights $w_I$. Write $W(u,v)$ for the total weight of the cuts in $\mathcal{F}$ that separate parties $u$ and $v$, and $r_x\coloneqq\min_{\ell\notin\{i,j\}}W(x,\ell)$ for $x\in\{i,j\}$. The Weighted Two-Party Cut Entanglement Inequality (WTPCEI) reads
\begin{equation}\label{intro:weighted-two-party-cut}
\min\{r_i,r_j\}\,\mathcal{E}_{A_iA_j}\leq\sum_{I\in\mathcal{F}}w_I\,\mathcal{E}_I,
\end{equation}
and its coefficient is optimal for every prescribed weighted family. It follows from a refinement in which the three mutual information terms of the tripartition of $A_i$, $A_j$, and the remaining parties enter with weights $W(i,j)$, $r_i$, and $r_j$. At fixed $\mathcal{E}_{A_i}$, $\mathcal{E}_{A_j}$, and $\mathcal{E}_{A_iA_j}$, this refinement gives the exact minimum of the weighted sum, attained by at most three independent bipartite pure states. The result recovers the EEI and the quadripartite triangle inequality, as well as the sharp coefficient $2(n-2)$ for the complete graph. For unit weights and $\{i,j\}\in\mathcal{F}$, it also shows that a polygon inequality with coefficient two holds if and only if every pair joining $\{i,j\}$ to the remaining parties is separated by another cut in $\mathcal{F}$. Applied to cyclic windows, it extends the EEI to every pair of parties, adjacent or not, with a mutual-information correction that grows with their cyclic separation. Applied to all $k$-party cuts, it bounds a two-party cut by the $k$-marginal total with the sharp coefficient $2\binom{n-2}{k-1}$ at every order $k$. Adjoining a party in a pure product state reduces a distinguished one-party cut to the same setting. The resulting Weighted One-Party Cut Entanglement Inequality (WOPCEI) reads $r_i\,\mathcal{E}_{A_i}\leq\sum_{I\in\mathcal{F}}w_I\,\mathcal{E}_I$ with the optimal coefficient $r_i\coloneqq\min_{\ell\neq i}W(i,\ell)$. Merging parties gives the same optimal coefficient, $\min_{u\in I_k,\,v\in\bar{I}_k}W(u,v)$, for a distinguished cut of any size whenever no cut of the family splits it or its complement. Together with complementarity, the optimal weighted coefficient is thereby determined for every distinguished cut when $n\leq5$.

We also obtain collective bounds for cuts containing a chosen party. Using Walecki decompositions of complete graphs into Hamiltonian cycles and, for even $n$, a perfect matching \cite{Alspach08}, we derive the Star Entanglement Inequality (SEI), which the following sharp bound supersedes for $n\geq5$. For $n\geq4$ and $2\leq k\leq\lfloor n/2\rfloor$, let $\mathscr{S}$ be a nonempty family of $q$ distinct $k$-party subsystems containing one fixed party: a partial $k$-uniform hyperstar. The Hyperstar Entanglement Inequality (HSEI) reads
\begin{equation}\label{intro:HSEI}
c_{n,k,q}\sum_{I_k\in\mathscr{S}}\mathcal{E}_{I_k}
\leq\mathcal{E}_{\mathfrak{I}_k}, \qquad
c_{n,k,q}\coloneqq\max\Bigl\{2,\frac{2}{q}\binom{n-2}{k-1}\Bigr\}.
\end{equation}
This coefficient is optimal uniformly over pure states, local dimensions, and partial hyperstars of size $q$. At marginal order two, the result gives a sharp collective star refinement for every prescribed number of selected incident edges.

Examples distinguish equality from strictness, including genuinely multipartite entangled states attaining HCEI equality. We characterize pure CEI equality states by adjacent pure bonds and obtain a dimension-independent trace-distance bound to EEI equality states. Applications give necessary bond-dimension bounds for matrix product states \cite{PVWC07,CPSV21}, constraints on correlations and squashed entanglement \cite{CW04}, and sequential-recovery error estimates \cite{FR15}.

Section~\ref{sec:preliminaries} fixes the entropy and entanglement notation and the graph and hypergraph conventions. Section~\ref{sec:Ent.Ineq.} develops the main entanglement inequalities, from the EPI to its generalized form, including average marginal bounds, a sharp $k$-marginal inequality, and optimal weighted bound for one- and two-party cut. It also derives graph and hypergraph inequalities for paths, cycles, parity-constrained families, stars, and hyperstars. Section~\ref{sec:applications} examines equality and near equality, tensor-network constraints, correlations, and recovery. Section~\ref{conclusion} summarizes the results and discusses open questions.

\section{Preliminaries}\label{sec:preliminaries}

\subsection{Von Neumann Entropy}\label{subsec:Von-Neumann_Entropy}

Let $\varrho$ be a density operator acting on a finite-dimensional Hilbert space $\mathcal{H}$. Its von Neumann entropy \cite{Neumann} is
\begin{equation}
\mathcal{S}(\varrho)\coloneqq-\operatorname{Tr}(\varrho\log_2\varrho).
\end{equation}
Throughout the paper, logarithms have base two, so entropy and entanglement entropy are measured in bits.

For a multipartite state, $\varrho_X$ denotes its reduced density operator on subsystem $X$. We recall the following standard entropy inequalities \cite{AL70,LR73}:
\begin{align}\label{NN}
\text{non-negativity:}\qquad
&\mathcal{S}(\varrho_A)\geq0,\\ \label{SA}
\text{subadditivity (SA):}\qquad
&\mathcal{S}(\varrho_{AB})
\leq\mathcal{S}(\varrho_A)+\mathcal{S}(\varrho_B),\\ \label{AL}
\text{Araki--Lieb inequality (AL):}\qquad
&\mathcal{S}(\varrho_{AB})
\geq\big|\mathcal{S}(\varrho_A)-\mathcal{S}(\varrho_B)\big|,\\ \label{WM}
\text{weak monotonicity (WM):}\qquad
&\mathcal{S}(\varrho_A)+\mathcal{S}(\varrho_C)
\leq\mathcal{S}(\varrho_{AB})+\mathcal{S}(\varrho_{BC}),\\ \label{SSA}
\text{strong subadditivity (SSA):}\qquad
&\mathcal{S}(\varrho_{ABC})+\mathcal{S}(\varrho_B)
\leq\mathcal{S}(\varrho_{AB})+\mathcal{S}(\varrho_{BC}).
\end{align}

\subsection{Entanglement Entropy}\label{subsec:Ent.Entropy}

Let $[n]\coloneqq\{1,\ldots,n\}$. For $1\leq k\leq n-1$, let
\begin{equation}\label{k-size-subsets}
\mathfrak{I}_k\coloneqq
\bigl\{\{j_1,\ldots,j_k\}\subset[n]\mid
j_1<\cdots<j_k\bigr\}
\end{equation}
denote the family of all distinct $k$-element subsets of $[n]$, each included exactly once. For each $I_k=\{j_1,\ldots,j_k\}\in\mathfrak{I}_k$, we write
$\bar{I}_k\coloneqq[n]\setminus I_k$ for its complement.
Whenever an ordering of the parties is required, we choose
a permutation $(i_1,\ldots,i_n)$ of $[n]$ such that
$I_k=\{i_1,\ldots,i_k\}$ and
$\bar{I}_k=\{i_{k+1},\ldots,i_n\}$.

Let $|\psi\rangle\in\medotimes_{j=1}^{n}\mathcal{H}_j \equiv\medotimes_{j=1}^{n}\mathbbm{C}^{d_j}$ be a pure $n$-partite quantum state with $n\geq2$. We label the $n$ parties by $A_1,\ldots,A_n$, where $A_j$ denotes the subsystem associated with the Hilbert space $\mathcal{H}_j$. For a subset $I_k=\{i_1,\ldots,i_k\}\subset[n]$, the corresponding composite subsystem is denoted by $A_{i_1}\cdots A_{i_k}$. The reduced density matrix of this subsystem is
\begin{equation}
\varrho_{I_k}=\mathrm{Tr}_{\bar{I}_k} \bigl(|\psi\rangle\langle\psi|\bigr).
\end{equation}

For the bipartition $I_k\,|\,\bar{I}_k$, the entanglement entropy is the von Neumann entropy of either reduced density operator,
\begin{equation}\label{complementarity}
\mathcal{E}_{I_k}\coloneqq
\mathsf{E}_{I_k\mid\bar{I}_k}(|\psi\rangle)
=\mathcal{S}(\varrho_{I_k})
=\mathcal{S}(\varrho_{\bar{I}_k})
=\mathcal{E}_{\bar{I}_k}.
\end{equation}
The equality of the complementary subsystem entropies follows from the purity of the global state. For a reduced mixed state $\varrho_{XY}$, we write $\mathsf{E}_{X\mid Y}(\varrho_{XY})$ for the entanglement between $X$ and $Y$, with the measure specified when needed. We omit the state argument when the state is clear from context.

In this paper, by \emph{$k$-marginal entanglement} (or \emph{$k$-party marginal entanglement}), we mean the entanglement between a $k$-party subsystem and its complement, i.e., the remainder of the $n$-partite system. Then, the total $k$-marginal entanglement can be written as
\begin{equation}
\mathcal{E}_{\mathfrak{I}_k}\coloneqq\sum_{I_k\in\mathfrak{I}_k}\mathcal{E}_{I_k}=\sum_{1\leq j_1<\cdots<j_k\leq n}\mathcal{E}_{A_{j_1}\cdots A_{j_k}}.
\end{equation}
When $n=2k$, complementary subsets are both included in this sum: they are distinct members of $\mathfrak{I}_k$, although their entanglement entropies are equal. We extend these conventions to the endpoint orders by setting $\mathfrak{I}_0\coloneqq\{\emptyset\}$, $\mathfrak{I}_n\coloneqq\{[n]\}$, and $\mathcal{E}_{\mathfrak{I}_0}=\mathcal{E}_{\mathfrak{I}_n}=0$, consistently with $\mathcal{E}_{\emptyset}=\mathcal{E}_{[n]}=0$. For $1\leq k\leq n-1$, define the average $k$-marginal entanglement by
\begin{equation}\label{average-k-marginal-entanglement}
\overline{\mathcal{E}}_k\coloneqq
\binom{n}{k}^{\!\!-1}\mathcal{E}_{\mathfrak{I}_k},
\end{equation}
and set $\overline{\mathcal{E}}_0=\overline{\mathcal{E}}_n=0$. Whenever convenient, we write $\mathcal{E}_{A_{i_1}\cdots A_{i_k}}$ instead of $\mathcal{E}_{I_k}$. In particular, $\mathcal{E}_{A_j}=\mathcal{S}(\varrho_{A_j})$ quantifies the entanglement between $A_j$ and its complementary subsystem, indexed by $[n]\setminus\{j\}$. Throughout, sums over empty index sets are understood to be zero.

\subsection{Graph and Hypergraph Terminology}
\label{subsec:graph}

We fix the graph and hypergraph conventions used throughout the paper; for general background, see Ref.~\cite{Berge73}. Vertices are identified with subsystem labels in $[n]$, so that an edge or hyperedge $I_k\in\mathfrak{I}_k$ specifies the bipartition $I_k\,|\,\bar{I}_k$ and its marginal entanglement $\mathcal{E}_{I_k}$.

A finite simple undirected graph is a pair $G=(V,E)$, where $V$ is a finite vertex set and $E\subseteq\binom{V}{2}$ is its edge set. The degree $\deg_G(v)$ counts the edges containing $v$; the graph is $r$-regular if every vertex has degree $r$. The complete graph $K_n$ has vertex set $[n]$ and edge set $E(K_n)=\{\{i,j\}\mid 1\leq i<j\leq n\}$, and is therefore $(n-1)$-regular.

A path is a sequence of distinct vertices with an edge between each consecutive pair. A cycle additionally joins the last vertex to the first and contains at least three vertices; $C_n$ denotes the cycle on $n$ vertices. A Hamiltonian cycle $\mathfrak{C}$ contains every vertex of the graph. A tree is a connected graph without cycles, and a spanning tree of $G$ is a tree subgraph containing every vertex of $G$. A matching consists of pairwise vertex-disjoint edges; it is perfect, or a $1$-factor, if it covers all vertices of the graph. A star centered at a vertex $v$ consists of edges joining $v$ to a set of other vertices. The full star at $v$ in $K_n$ contains all $n-1$ edges incident with $v$.

An edge decomposition of $G$ into subgraphs $G_1,\ldots,G_t$ means that their edge sets partition $E(G)$. We write $G=G_1\sqcup\cdots\sqcup G_t$, where the subgraphs may share vertices. A Hamiltonian decomposition consists entirely of Hamiltonian cycles. We use the classical Walecki construction \cite{Alspach08}, which gives
\begin{equation}\label{Walecki-odd}
K_{2r+1}=\mathfrak{C}^{(1)}\sqcup\cdots\sqcup\mathfrak{C}^{(r)}
\end{equation}
for odd order, and
\begin{equation}\label{Walecki-even}
K_{2r}=M\sqcup\mathfrak{C}^{(1)}\sqcup\cdots\sqcup\mathfrak{C}^{(r-1)}
\end{equation}
for even order. Here the $\mathfrak{C}^{(s)}$ are Hamiltonian cycles and $M$ is a perfect matching; equivalently, $K_{2r}-M$ admits a Hamiltonian decomposition.

A finite hypergraph without repeated hyperedges is a pair $\mathscr{H}=(V,\mathscr{F})$, where $V$ is a finite vertex set and $\mathscr{F}\subseteq 2^V\setminus\{\emptyset\}$ is a family of nonempty subsets of $V$, called hyperedges. Containment between distinct hyperedges is allowed. The degree of $v\in V$ is
\begin{equation*}
\deg_{\mathscr{H}}(v)\coloneqq\bigl|\{F\in\mathscr{F}\mid v\in F\}\bigr|.
\end{equation*}
The hypergraph is $r$-regular if every vertex has degree $r$, and $k$-uniform if every hyperedge has cardinality $k$. The complete $k$-uniform hypergraph $\mathscr{K}_n^{(k)}$ on $[n]$ has hyperedge set
\begin{equation*}
\mathscr{F}\bigl(\mathscr{K}_n^{(k)}\bigr)=\mathfrak{I}_k.
\end{equation*}
In particular, $\mathscr{K}_n^{(2)}=K_n$.

For $v\in[n]$, the full $k$-uniform hyperstar centered at $v$ is
\begin{equation}
\mathscr{S}_{v}^{(k)}\coloneqq\{I_k\in\mathfrak{I}_k\mid v\in I_k\},
\qquad \bigl|\mathscr{S}_{v}^{(k)}\bigr|=\binom{n-1}{k-1}.
\end{equation}
We call any subfamily $\mathscr{S}\subseteq\mathscr{S}_{v}^{(k)}$ a partial hyperstar. Its hyperedges may also intersect outside the center $v$.

We also consider ordered families of consecutive hyperedges. For $1\leq m\leq n-2$, the sets
\begin{equation}\label{I_m+1-subsets}
I_{m+1}^{(j)}=\{j,j+1,\ldots,j+m\}, \qquad 1\leq j\leq n-m,
\end{equation}
form what we call an $(m+1)$-uniform hyperchain. Consecutive hyperedges overlap in exactly $m$ vertices. Using the same definition with indices interpreted cyclically modulo $n$, the family $\{I_{m+1}^{(j)}\}_{j\in[n]}$ is called the corresponding hypercycle. For $m=1$, these constructions recover the path and cycle on $n$ vertices.

Finally, the symmetric difference of two subsets $I,J\subseteq[n]$ is
\begin{equation}
I\triangle J\coloneqq(I\setminus J)\cup(J\setminus I).
\end{equation}
For example, $\{1,2,4\}\triangle\{2,3,4\}=\{1,3\}$. We write $\bigtriangleup_{t=1}^{q}I^{(t)}$ for the iterated symmetric difference of several subsets. If the members of the family are regarded as hyperedges, a vertex belongs to this set precisely when it occurs in an odd number of the subsets. Thus, for a hypergraph $\mathscr{H}=(V,\mathscr{F})$,
\begin{equation}
\bigtriangleup_{F\in\mathscr{F}}F =\{v\in V\mid \deg_{\mathscr{H}}(v)\ \text{is odd}\}.
\end{equation}
Consequently, the total symmetric difference is $\emptyset$ when all vertex degrees are even and $V$ when all are odd. This incidence-parity interpretation underlies the graph- and hypergraph-theoretic entanglement polygon inequalities developed below.

\section{Entanglement Inequalities}\label{sec:Ent.Ineq.}

\subsection{Polygon and Marginal Inequalities}\label{subsec:Polygon_Marginal}

We begin by recalling the entanglement polygon inequality (EPI), which constrains the one-versus-rest entanglement of a pure multipartite state. Polygon inequalities were first obtained for the spectra of one-qubit reduced states in \cite{HSS03}, and were subsequently formulated in terms of entanglement measures for multipartite qubit systems \cite{QAE18} and arbitrary-dimensional multipartite systems \cite{YYL22}.

\begin{theorem}[Entanglement Polygon Inequality \cite{HSS03,QAE18,YYL22}]\label{Theorem:EPI}
Let $|\psi\rangle\in\medotimes_{j=1}^{n}\mathbbm{C}^{d_j}$ be a pure $n$-partite quantum state with $n\geq2$. Then, the one-marginal entanglement values satisfy the polygon inequality. That is, for all $k\in[n]$,
\begin{equation}\label{EPI}
2\,\mathcal{E}_{A_k}\leq\sum_{j=1}^{n}\mathcal{E}_{A_j}.
\end{equation}
\end{theorem}
\begin{proof}
Fix $k\in[n]$ and write $[n]\setminus\{k\}=\{i_1,\ldots,i_{n-1}\}$. By purity of the global state,
\begin{equation*}
\mathcal{E}_{A_k}=\mathsf{E}_{A_k|A_{i_1}\cdots A_{i_{n-1}}}
(|\psi\rangle)=\mathcal{S}(\varrho_{A_k})=\mathcal{S}(\varrho_{A_{i_1}\cdots A_{i_{n-1}}}).
\end{equation*}
Repeated application of SA therefore gives
\begin{equation*}
\mathcal{E}_{A_k}=\mathcal{S}(\varrho_{A_{i_1}\cdots A_{i_{n-1}}})
\leq\mathcal{S}(\varrho_{A_{i_1}})+\mathcal{S}(\varrho_{A_{i_2}\cdots A_{i_{n-1}}})\leq\cdots\leq\sum_{l=1}^{n-1}\mathcal{S}(\varrho_{A_{i_l}})
=\sum_{j\neq k}\mathcal{E}_{A_j}.
\end{equation*}
Adding $\mathcal{E}_{A_k}$ to both sides yields Eq.~\eqref{EPI}.
\end{proof}

The EPI concerns only one-versus-rest bipartitions.
More generally, Ref.~\cite{YYL22} established one-sided
distribution inequalities for arbitrary multipartite
bipartitions, first for $q$-concurrence and subsequently
for unified-$(r,s)$ entropy entanglement. In the present
von Neumann entropy setting, applying SA to
both sides of a bipartition gives a bound in terms of the
total one-marginal entanglement.

\begin{theorem}[Marginal Entanglement Inequality \cite{YYL22}]
\label{Theorem:MEI}
Let $|\psi\rangle\in\medotimes_{j=1}^{n}\mathbbm{C}^{d_j}$ be a pure $n$-partite quantum state with $n\geq2$. For every $1\leq k\leq n-1$ and every $I_k\in\mathfrak{I}_k$, the $k$-marginal entanglement is bounded above by one half of the total one-marginal entanglement:
\begin{equation}\label{eq-MEI}
2\,\mathcal{E}_{I_k}
\leq\mathcal{E}_{\mathfrak{I}_1}.
\end{equation}
\end{theorem}
\begin{proof}
Choose an ordering $(i_1,\ldots,i_n)$ of $[n]$ such that
$I_k=\{i_1,\ldots,i_k\}$ and
$\bar{I}_k=\{i_{k+1},\ldots,i_n\}$.
Repeated SA gives
\begin{equation*}
\mathcal{E}_{I_k}=\mathcal{E}_{A_{i_1}\cdots A_{i_k}}
=\mathcal{S}(\varrho_{A_{i_1}\cdots A_{i_k}})
\leq \mathcal{S}(\varrho_{A_{i_1}})+\mathcal{S}(\varrho_{A_{i_2}\cdots A_{i_k}}) \leq \cdots \leq
\sum_{j=1}^{k}\mathcal{S}(\varrho_{A_{i_j}})=\sum_{j\in I_k}\mathcal{E}_{A_j}.
\end{equation*}
Applying the same argument to the complementary subsystem gives
\begin{equation*}
\mathcal{E}_{\bar{I}_k}=\mathcal{E}_{A_{i_{k+1}}\cdots A_{i_n}}
=\mathcal{S}(\varrho_{A_{i_{k+1}}\cdots A_{i_n}})
\leq \mathcal{S}(\varrho_{A_{i_{k+1}}}) + \mathcal{S}(\varrho_{A_{i_{k+2}}\cdots A_{i_n}}) \leq \cdots \leq
\sum_{j=k+1}^{n}\mathcal{S}(\varrho_{A_{i_j}})=\sum_{j\in\bar{I}_k}\mathcal{E}_{A_j}.
\end{equation*}
Either estimate holds with equality when its subsystem
contains only one party. By pure-state complementarity,
$\mathcal{E}_{I_k}=\mathcal{E}_{\bar{I}_k}$.
Adding the two estimates therefore yields
\begin{equation*}
2\,\mathcal{E}_{I_k}
\leq
\sum_{j\in I_k}\mathcal{E}_{A_j}
+\sum_{j\in\bar{I}_k}\mathcal{E}_{A_j}
=\mathcal{E}_{\mathfrak{I}_1}.
\end{equation*}
\end{proof}

\begin{remark}
For $k=1$, the MEI in Theorem~\ref{Theorem:MEI}
reduces to the EPI in Theorem~\ref{Theorem:EPI}.
Thus, the MEI extends the polygon inequality from
one-party cuts to arbitrary subsystem bipartitions.
\end{remark}

The proof of Theorem~\ref{Theorem:MEI} gives a stronger
bound when its two subadditivity estimates are retained
separately.

\begin{theorem}[Strong Marginal Entanglement Inequality]
\label{Theorem:SMEI}
Let $|\psi\rangle\in
\medotimes_{j=1}^{n}\mathbbm{C}^{d_j}$ be a pure
$n$-partite quantum state with $n\geq2$. For every
$1\leq k\leq n-1$ and $I_k\in\mathfrak{I}_k$,
\begin{equation}\label{SMEI}
\mathcal{E}_{I_k}\leq\min\Big\{\sum_{j\in I_k}\mathcal{E}_{A_j},\sum_{j\in\bar{I}_k}\mathcal{E}_{A_j}\Big\}.
\end{equation}
\end{theorem}
\begin{proof}
The proof of Theorem~\ref{Theorem:MEI} establishes
$\mathcal{E}_{I_k}\leq
\sum_{j\in I_k}\mathcal{E}_{A_j}$ and
$\mathcal{E}_{\bar{I}_k}\leq
\sum_{j\in\bar{I}_k}\mathcal{E}_{A_j}$.
Since $\mathcal{E}_{I_k}=\mathcal{E}_{\bar{I}_k}$,
taking the minimum gives Eq.~\eqref{SMEI}.
\end{proof}

\begin{remark}\label{rem:subadditivity-EPI-MEI-SMEI}
The EPI, MEI, and SMEI compare entanglement across bipartitions of the globally pure state. Consequently, they remain valid for any bipartite entanglement measure that agrees with the von Neumann entropy of entanglement on every pure bipartite state: all quantities occurring in these inequalities then have exactly the same values. More generally, their proofs apply when the pure-state measure is induced by a functional $F$ satisfying $F(\varrho_I)=F(\varrho_{\bar{I}})$ for complementary reductions and subadditivity on the mixed marginals used in the proofs.
\end{remark}

Since Theorem~\ref{Theorem:SMEI} applies to every size-$k$ subset $I_k\in\mathfrak I_k$, averaging over all $\binom{n}{k}$ such subsets yields a sharper upper bound than that obtained by averaging the bipartition-independent MEI.

\begin{corollary}[Average $k$-Marginal Entanglement Inequality]
\label{cor:average-k-marginal-entanglement}
Let $|\psi\rangle\in\medotimes_{j=1}^{n}\mathbbm{C}^{d_j}$ be a pure $n$-partite quantum state with $n\geq2$. For every $1\leq k\leq n-1$,
\begin{equation}\label{k-marginal-1-marginal}
\overline{\mathcal{E}}_k\leq
\frac{\min\{k,n-k\}}{n}\,\mathcal{E}_{\mathfrak{I}_1}.
\end{equation}
\end{corollary}
\begin{proof}
By Theorem~\ref{Theorem:SMEI}, $\mathcal{E}_{I_k}\leq\sum_{j\in I_k}\mathcal{E}_{A_j}$ for every $I_k\in\mathfrak{I}_k$. Summing over all such subsets $I_k\in\mathfrak{I}_k$ and dividing by $\binom{n}{k}$ gives
\begin{equation*}
\binom{n}{k}^{\!\!-1}\sum_{I_k\in\mathfrak{I}_k}\mathcal{E}_{I_k}
\leq \frac{k}{n}\sum_{j=1}^{n}\mathcal{E}_{A_j},
\end{equation*}
since each single-party term occurs $\binom{n-1}{k-1}$ times. Applying the same argument to the complementary subsets $\bar{I}_k$ and using $\mathcal{E}_{I_k}=\mathcal{E}_{\bar{I}_k}$ gives the corresponding bound with coefficient $(n-k)/n$. Taking the minimum of these two bounds completes the proof.
\end{proof}

In particular, Eq.~\eqref{k-marginal-1-marginal} implies the upper bound $\mathcal{E}_{\mathfrak{I}_1}/2$, which also follows by averaging Theorem~\ref{Theorem:MEI}, and improves the coefficient whenever $k\neq n/2$. The refined bound is sharp whenever an absolutely maximally entangled state $\mathrm{AME}(n,d)$ exists. Such a state is a pure state of $n$ parties of local dimension $d$ whose marginals on at most $\lfloor n/2\rfloor$ parties are maximally mixed \cite{RBRLZ26}. Consequently, by purity, every $k$-party marginal has entropy $\min\{k,n-k\}\log_2 d$, while every one-party marginal has entropy $\log_2 d$, so the bound is saturated.

A further consequence of the SMEI is obtained by taking powers of the marginal entanglement. For $0<\alpha\leq1$, the function $x\mapsto x^\alpha$ is monotonically increasing and subadditive on $\mathbbm{R}_{\geq0}$, i.e.,
\begin{equation*}
\Bigl(\sum_i x_i\Bigr)^{\!\alpha}\leq\sum_i x_i^\alpha.
\end{equation*}
Applying this property separately to the two complementary bounds in Theorem~\ref{Theorem:SMEI} gives the following family of inequalities.

\begin{proposition}\label{prop:SMEI-alpha}
Let $|\psi\rangle\in\medotimes_{j=1}^{n}\mathbbm{C}^{d_j}$ be a pure $n$-partite state with $n\geq2$, and let $\alpha\in(0,1]$. For every $I_k\in\mathfrak{I}_k$,
\begin{equation}\label{alpha-power-SMEI}
2\,\mathcal{E}_{I_k}^\alpha
\leq \sum_{j=1}^{n}\mathcal{E}_{A_j}^\alpha.
\end{equation}
In particular, for $k=1$ this reduces to the power form of the EPI:
for every $l\in[n]$,
\begin{equation}
2\,\mathcal{E}_{A_l}^\alpha\leq\sum_{j=1}^{n}\mathcal{E}_{A_j}^\alpha.
\end{equation}
\end{proposition}
\begin{proof}
By Theorem~\ref{Theorem:SMEI}, $\mathcal{E}_{I_k}\leq\sum_{j\in I_k}\mathcal{E}_{A_j}$ and $\mathcal{E}_{I_k}\leq\sum_{j\in\bar{I}_k}\mathcal{E}_{A_j}$. Since $f(x)=x^\alpha$ is a monotonically increasing function for $\alpha>0$, raising both inequalities to the power $\alpha$ gives
\begin{equation*}
\mathcal{E}_{I_k}^{\alpha}
\leq\Bigl(\sum_{j\in I_k}\mathcal{E}_{A_j}\Bigr)^{\!\alpha},
\qquad \qquad
\mathcal{E}_{I_k}^{\alpha}\leq\Bigl(\sum_{j\in\bar{I}_k}\mathcal{E}_{A_j}\Bigr)^{\!\alpha}.
\end{equation*}
For $0<\alpha\leq1$, subadditivity of $x^\alpha$ then implies $\mathcal{E}_{I_k}^{\alpha}\leq\sum_{j\in I_k}\mathcal{E}_{A_j}^{\alpha}$ and $\mathcal{E}_{I_k}^{\alpha}\leq\sum_{j\in\bar{I}_k}\mathcal{E}_{A_j}^{\alpha}$. Adding the two inequalities yields Eq.~\eqref{alpha-power-SMEI}. The power form of the EPI follows by taking $k=1$ and $I_1=\{l\}$, for any $l\in[n]$.
\end{proof}

\subsection{Chain, Edge, and Higher-Order Inequalities}\label{subsec:Chain_Edge}

We now incorporate an ordering of the subsystems and derive entanglement constraints associated with a path and a cycle. The key ingredient is the SSA of the von Neumann entropy (see Eq.~\eqref{SSA}). For three subsystems $X,Y,Z$, we recall that the conditional mutual information is defined by
\begin{equation}
\mathcal{I}(X:Z\mid Y)\coloneqq
\mathcal{S}(\varrho_{XY})+\mathcal{S}(\varrho_{YZ})
-\mathcal{S}(\varrho_Y)-\mathcal{S}(\varrho_{XYZ}),
\end{equation}
and SSA is equivalent to $\mathcal{I}(X:Z\mid Y)\geq0$. With trivial conditioning subsystem, $\mathcal{I}(X:Z)\coloneqq\mathcal{S}(\varrho_X)+\mathcal{S}(\varrho_Z)-\mathcal{S}(\varrho_{XZ})$ is the mutual information.

The following result gives not only the chain inequality itself, but also an exact expression for its deficit in terms of conditional mutual information.

\begin{theorem}[Chain Entanglement Inequality]\label{Theorem:CEI}
Let $|\psi\rangle\in\medotimes_{j=1}^{n}\mathbbm{C}^{d_j}$ be a pure $n$-partite quantum state with $n\geq3$. Then
\begin{equation}\label{CEI-deficit}
\sum_{j=1}^{n-1}\mathcal{E}_{A_jA_{j+1}}-\sum_{j=1}^{n-2}\mathcal{E}_{A_{j+1}}=
\sum_{j=1}^{n-2}\mathcal{I}\big(A_1\cdots A_j:A_{j+2}\mid A_{j+1}\big).
\end{equation}
Consequently,
\begin{equation}\label{CEI}
\sum_{j=1}^{n-2}\mathcal{E}_{A_{j+1}} \leq
\sum_{j=1}^{n-1}\mathcal{E}_{A_jA_{j+1}}.
\end{equation}
Moreover, equality holds if and only if
\begin{equation}\label{Markov-chain}
\mathcal{I}\big(A_1\cdots A_j:A_{j+2}\mid A_{j+1}\big)=0,
\qquad \forall~1\leq j\leq n-2.
\end{equation}
\end{theorem}
\begin{proof}
For each $1\leq j\leq n-2$, the definition of conditional mutual information gives
\begin{equation*}
\mathcal{I}\big(A_1\cdots A_j:A_{j+2}\mid A_{j+1}\big) =
\mathcal{S}(\varrho_{A_1\cdots A_{j+1}})+\mathcal{S}(\varrho_{A_{j+1}A_{j+2}})
-\mathcal{S}(\varrho_{A_{j+1}})-\mathcal{S}(\varrho_{A_1\cdots A_{j+2}}).
\end{equation*}
Summing over $j=1,\ldots,n-2$, the prefix entropies telescope:
\begin{align*}
\sum_{j=1}^{n-2}\mathcal{I}\big(A_1\cdots A_j:A_{j+2}\mid A_{j+1}\big)
&=\mathcal{S}(\varrho_{A_1A_2})
+\sum_{j=1}^{n-2}\mathcal{S}(\varrho_{A_{j+1}A_{j+2}})
-\sum_{j=1}^{n-2}\mathcal{S}(\varrho_{A_{j+1}})
-\mathcal{S}(\varrho_{A_1\cdots A_n}) \\
&=\sum_{j=1}^{n-1}\mathcal{S}(\varrho_{A_jA_{j+1}})
-\sum_{j=1}^{n-2}\mathcal{S}(\varrho_{A_{j+1}})
-\mathcal{S}(\varrho_{A_1\cdots A_n}).
\end{align*}
Since the global state is pure, $\mathcal{S}(\varrho_{A_1\cdots A_n})=0$, and the remaining entropy terms are precisely the corresponding marginal entanglement. This proves Eq.~\eqref{CEI-deficit}. By SSA, every conditional mutual information on the right-hand side of Eq.~\eqref{CEI-deficit} is nonnegative for every $1\leq j\leq n-2$. Hence Eq.~\eqref{CEI-deficit} immediately implies Eq.~\eqref{CEI}. Moreover, equality in Eq.~\eqref{CEI} holds if and only if every term on the right-hand side of Eq.~\eqref{CEI-deficit} vanishes, which gives Eq.~\eqref{Markov-chain}.
\end{proof}

\begin{remark}\label{rem:CEI-Markov}
The equality conditions in Eq.~\eqref{Markov-chain} involve growing prefixes of the ordered path. For $n\geq4$, the local conditions $\mathcal{I}(A_j:A_{j+2}\mid A_{j+1})=0$ alone are not sufficient to saturate the CEI. The stronger prefix conditions are essential; an explicit separation is given in Sec.~\ref{sec:applications}. Their structural interpretation is discussed there together with the pure-state equality characterization.
\end{remark}

\begin{remark}\label{rem:mixed-chain}
The purity assumption in Theorem~\ref{Theorem:CEI} is required only to reinterpret the subsystem entropies as marginal entanglement. Indeed, the telescoping argument holds for an arbitrary, possibly mixed, $n$-partite state and gives
\begin{equation}\label{mixed-chain-entropy}
\mathcal{S}(\varrho_{A_1\cdots A_n})
+
\sum_{j=1}^{n-2}\mathcal{S}(\varrho_{A_{j+1}})
\leq
\sum_{j=1}^{n-1}\mathcal{S}(\varrho_{A_jA_{j+1}}),
\end{equation}
with exact deficit $\sum_{j=1}^{n-2}\mathcal{I}(A_1\cdots A_j:A_{j+2}\mid A_{j+1})$. Global purity sets $\mathcal{S}(\varrho_{A_1\cdots A_n})=0$ and identifies the remaining subsystem entropies with the corresponding subsystem-versus-rest entanglement, giving the CEI. The relation of this mixed-state bound and its higher-order versions to entropy decompositions is discussed in Sec.~\ref{sec:applications}.
\end{remark}

We next close the path into a cycle. Each edge of the resulting cycle corresponds to the nearest-neighbors-versus-rest entanglement $\mathcal{E}_{A_kA_{k+1}}$, namely, the entanglement across the bipartition $A_kA_{k+1}\,|\,\overline{A_kA_{k+1}}$, which is a special case of two-marginal entanglement.

\begin{theorem}[Edge Entanglement Inequality]\label{Theorem:EEI}
Let $|\psi\rangle\in\medotimes_{j=1}^{n}\mathbbm{C}^{d_j}$ be a pure $n$-partite quantum state with $n\geq3$, and adopt the cyclic convention $A_{j+n}\equiv A_j$. Then, for every $k\in[n]$,
\begin{equation}\label{EEI}
2\,\mathcal{E}_{A_kA_{k+1}}\leq\sum_{j=1}^{n}\mathcal{E}_{A_jA_{j+1}}.
\end{equation}
\end{theorem}
\begin{proof}
It is sufficient, by cyclic relabeling, to prove the result for the edge $A_n\text{---}A_1$. By purity of the global state, $\mathcal{E}_{A_1A_n}=\mathcal{S}(\varrho_{A_1A_n})=\mathcal{S}(\varrho_{A_2\cdots A_{n-1}})$. Repeated application of SA gives
\begin{equation*}
\mathcal{E}_{A_1A_n}=\mathcal{S}(\varrho_{A_2\cdots A_{n-1}})
\leq \mathcal{S}(\varrho_{A_2})+\mathcal{S}(\varrho_{A_3\cdots A_{n-1}}) \leq \cdots \leq \sum_{j=2}^{n-1}\mathcal{S}(\varrho_{A_j})
=\sum_{j=1}^{n-2}\mathcal{E}_{A_{j+1}}.
\end{equation*}
Applying the CEI in Eq.~\eqref{CEI} and adding $\mathcal{E}_{A_1A_n}$ to both sides yields
\begin{equation*}
2\,\mathcal{E}_{A_1A_n}
\leq
\sum_{j=1}^{n}\mathcal{E}_{A_jA_{j+1}},
\end{equation*}
where $A_{n+1}\equiv A_1$. Cyclic relabeling gives the result for every edge of the cycle.
\end{proof}

The slack in the EEI also admits an exact information-theoretic decomposition. Define the CEI deficit along the path $A_1\text{---}\cdots\text{---}A_n$ by
\begin{equation}
\Delta_{\mathrm{CEI}}\coloneqq
\sum_{j=1}^{n-1}\mathcal{E}_{A_jA_{j+1}}-\sum_{j=2}^{n-1}\mathcal{E}_{A_j},
\end{equation}
and define the total correlation among the interior subsystems by
\begin{equation}
\mathcal{T}(A_2:\cdots:A_{n-1})\coloneqq
\sum_{j=2}^{n-1}\mathcal{S}(\varrho_{A_j}) -
\mathcal{S}(\varrho_{A_2\cdots A_{n-1}}).
\end{equation}
Repeated SA implies $\mathcal{T}(A_2:\cdots:A_{n-1})\geq0$. Using purity, $\mathcal{S}(\varrho_{A_2\cdots A_{n-1}})=\mathcal{E}_{A_1A_n}$, and hence
\begin{equation}\label{EEI-deficit}
\sum_{j=1}^{n}\mathcal{E}_{A_jA_{j+1}}-2\,\mathcal{E}_{A_1A_n}=\Delta_{\mathrm{CEI}}+\mathcal{T}(A_2:\cdots:A_{n-1}).
\end{equation}
Thus, the EEI deficit contains two nonnegative contributions: the chain deficit along the complementary path and the total correlation among its interior subsystems.

\begin{remark}
For $n=3$, both the CEI and the EEI reduce to the EPI. Indeed, for $\{i,j,k\}=[3]$ purity gives $\mathcal{E}_{A_iA_j}=\mathcal{E}_{A_k}$, so both inequalities become the usual triangle inequality among the three one-marginal entanglement.
\end{remark}

\begin{remark}\label{Rem.n=4}
For $n=4$, the EEI becomes trivial because complementary two-party marginals have equal entropy. For example, $\mathcal{E}_{AB}=\mathcal{E}_{CD}$ and $\mathcal{E}_{BC}=\mathcal{E}_{AD}$, and therefore the EEI for the cycle $A\text{---}B\text{---}C\text{---}D\text{---}A$ reduces to the non-negativity of the remaining two-marginal entanglement. In contrast, the CEI remains nontrivial. For the path $A\text{---}B\text{---}C\text{---}D$, Eq.~\eqref{CEI-deficit} becomes
\begin{equation}
\mathcal{E}_{AB}+\mathcal{E}_{BC}+\mathcal{E}_{CD}-\mathcal{E}_{B}-\mathcal{E}_{C}=\mathcal{I}(A:C\mid B)+\mathcal{I}(AB:D\mid C)\geq0.
\end{equation}
\end{remark}

For four parties, one can furthermore derive a stronger polygon relation among the three inequivalent balanced bipartitions.

\begin{corollary}[Quadripartite Two-Marginal Triangle Inequality]\label{cor:four-party-triangle}
Let $|\psi\rangle_{ABCD}\in\medotimes_{j=1}^{4}\mathbbm{C}^{d_j}$ be a pure quadripartite quantum state. Then
\begin{equation}\label{four-party-triangle}
\mathcal{E}_{AB}\leq\mathcal{E}_{AC}+\mathcal{E}_{AD},
\end{equation}
together with the two inequalities obtained by permuting the three balanced bipartitions $AB|CD$, $AC|BD$, and $AD|BC$.
\end{corollary}
\begin{proof}
SA gives $\mathcal{E}_{AB}=\mathcal{S}(\varrho_{AB})\leq\mathcal{S}(\varrho_A)+\mathcal{S}(\varrho_B)$. WM applied to the tripartite marginal $\varrho_{ABC}$ yields $\mathcal{S}(\varrho_A)+\mathcal{S}(\varrho_B)\leq\mathcal{S}(\varrho_{AC})+\mathcal{S}(\varrho_{BC})$. Finally, purity implies $\mathcal{S}(\varrho_{BC})=\mathcal{S}(\varrho_{AD})$, and hence $\mathcal{E}_{AB}\leq\mathcal{E}_{AC}+\mathcal{E}_{AD}$. The remaining inequalities follow by permutation of the parties.
\end{proof}

A natural extension of the EEI is obtained by allowing all two-marginal entanglement $\mathcal{E}_{A_iA_j}$, rather than restricting to the nearest-neighbors-versus-rest entanglement $\mathcal{E}_{A_kA_{k+1}}$ associated with the edges of a fixed cycle. By representing each two-marginal entanglement $\mathcal{E}_{A_iA_j}$ as an edge of the complete graph $K_n$, the closed chain in Theorem~\ref{Theorem:EEI} corresponds to a Hamiltonian cycle in $K_n$. Enlarging from the edge set of such a cycle to the full edge set of $K_n$ leads naturally to an inequality involving all two-marginal entanglement. As we show below, this inequality retains the polygon structure of the EPI in Theorem~\ref{Theorem:EPI}, thereby extending the entanglement polygon inequality from one-marginal to two-marginal entanglement.

\begin{corollary}[Two-Marginal Entanglement Polygon Inequality]\label{cor:2-EPI}
Let $|\psi\rangle\in\medotimes_{j=1}^{n}\mathbbm{C}^{d_j}$ be a pure $n$-partite quantum state with $n\geq3$. Then, for every pair of distinct indices $k,l\in[n]$,
\begin{equation}\label{2-EPI}
2\,\mathcal{E}_{A_kA_l}\leq\sum_{1\leq i<j\leq n}\mathcal{E}_{A_iA_j}.
\end{equation}
\end{corollary}
\begin{proof}
Since $K_n$ is complete, there exists a Hamiltonian cycle containing the edge $\{k,l\}$. Applying the EEI in Theorem~\ref{Theorem:EEI} to this cycle gives
\begin{equation*}
2\,\mathcal{E}_{A_kA_l} \leq
\sum_{\{i,j\}\in E(\mathfrak{C})}\mathcal{E}_{A_iA_j},
\end{equation*}
where $E(\mathfrak{C})$ denotes the edge set of the chosen Hamiltonian cycle. Since all marginal entanglement values are nonnegative and $E(\mathfrak{C})\subseteq E(K_n)$,
\begin{equation*}
\sum_{\{i,j\}\in E(\mathfrak{C})}\mathcal{E}_{A_iA_j}\leq
\sum_{1\leq i<j\leq n}\mathcal{E}_{A_iA_j}.
\end{equation*}
Combining the two inequalities proves Eq.~\eqref{2-EPI}.
\end{proof}

\begin{remark}
For $n=3$, the TMEPI reduces to the EPI by complementarity. For $n=4$, Eq.~\eqref{2-EPI} is weaker than Corollary~\ref{cor:four-party-triangle}, because the six two-marginal terms occur in three complementary pairs. For larger systems it is an all-edge consequence of the cycle inequality, and is in turn implied by the sharp higher-order marginal bound below.
\end{remark}

\begin{remark}\label{rem:SSA-generality}
The direct proof of the CEI uses SSA. The derivation of the EEI additionally uses SA and complementarity for pure states, while the passage from the EEI to the TMEPI uses only non-negativity of the marginal entanglement. Consequently, analogous inequalities can be obtained from other entropy functionals whenever the corresponding forms of SSA, SA, and pure-state complementarity hold. In particular, replacing the von Neumann entropy by the linear entropy $\mathcal{S}_L(\varrho)=1-\operatorname{Tr}(\varrho^2)$ does not preserve this derivation, since the linear entropy does not satisfy SSA in general \cite{PV15}.
\end{remark}

The EPI and the TMEPI establish polygon inequalities for one- and two-marginal entanglement, respectively. To extend this structure to arbitrary marginal order, we compare the total entanglement carried by marginals of successive orders and establish a hierarchy for the corresponding average marginal entanglement.

\begin{lemma}[Average Marginal Entanglement Hierarchy]
\label{lem:average-marginal-hierarchy}
Let $|\psi\rangle\in\medotimes_{j=1}^{n}\mathbbm{C}^{d_j}$ be a pure $n$-partite quantum state. For every $k$ with $1\leq k<\lfloor n/2\rfloor$, the average $k$-marginal entanglement is bounded above by the average $(k+1)$-marginal entanglement, that is,
\begin{equation}\label{average-marginal-concave}
\overline{\mathcal{E}}_{k}\leq\overline{\mathcal{E}}_{k+1}.
\end{equation}
Consequently, for $n\geq4$,
\begin{equation}\label{average-marginal-hierarchy}
\overline{\mathcal{E}}_1\leq\overline{\mathcal{E}}_2
\leq\cdots\leq\overline{\mathcal{E}}_{\lfloor n/2\rfloor}.
\end{equation}
\end{lemma}
\begin{proof}
By Eq.~\eqref{complementarity}, the map $I_k\mapsto\bar{I}_k$ gives a bijection between $\mathfrak{I}_k$ and $\mathfrak{I}_{n-k}$, and hence $\overline{\mathcal{E}}_{k}=\overline{\mathcal{E}}_{n-k}$. Fix $1\leq r\leq n-1$, $J_{r-1}\in\mathfrak{I}_{r-1}$, and distinct $a,b\in[n]\setminus J_{r-1}$. Applying SSA to the subsystem entropies gives
\begin{equation*}
\mathcal{E}_{J_{r-1}\cup\{a,b\}}+\mathcal{E}_{J_{r-1}}\leq
\mathcal{E}_{J_{r-1}\cup\{a\}}+\mathcal{E}_{J_{r-1}\cup\{b\}}.
\end{equation*}
Sum over all $J_{r-1}$ and all unordered pairs $\{a,b\}$ in their complements. Each $(r+1)$-marginal term occurs $\binom{r+1}{2}$ times, each $(r-1)$-marginal term occurs $\binom{n-r+1}{2}$ times, and each $r$-marginal term occurs $r(n-r)$ times. Thus
\begin{equation*}
\binom{r+1}{2}\mathcal{E}_{\mathfrak{I}_{r+1}}
+\binom{n-r+1}{2}\mathcal{E}_{\mathfrak{I}_{r-1}}
\leq r(n-r)\mathcal{E}_{\mathfrak{I}_r}.
\end{equation*}
Since
\begin{equation*}
\binom{n-r+1}{2}\binom{n}{r-1}=\binom{r+1}{2}\binom{n}{r+1}
=\frac{r(n-r)}{2}\binom{n}{r},
\end{equation*}
dividing by this common factor gives $\overline{\mathcal{E}}_{r+1}-\overline{\mathcal{E}}_r \leq\overline{\mathcal{E}}_r-\overline{\mathcal{E}}_{r-1}$. The successive differences $\delta_r\coloneqq\overline{\mathcal{E}}_r-\overline{\mathcal{E}}_{r-1}$ are therefore nonincreasing. By symmetry, $\delta_{n+1-r}=-\delta_r$. For $1\leq k<\lfloor n/2\rfloor$, we have $k+1\leq n-k$, so $\delta_{k+1}\geq\delta_{n-k}=-\delta_{k+1}$. Thus $\overline{\mathcal{E}}_k\leq\overline{\mathcal{E}}_{k+1}$, which gives Eq.~\eqref{average-marginal-hierarchy} for $n\geq4$.
\end{proof}

\begin{remark}
By complementarity, the average marginal entanglement profile $(\overline{\mathcal{E}}_k)_{k=0}^{n}$ is symmetric. The proof above also shows that it is discretely concave. For every $k$ with $1\leq k\leq n-1$,
\begin{equation}\label{average-marginal-concavity}
\overline{\mathcal{E}}_{k-1}+\overline{\mathcal{E}}_{k+1}
\leq 2\,\overline{\mathcal{E}}_k.
\end{equation}
This inequality is known: it is the symmetrized form of strong subadditivity obtained in Pippenger's characterization of symmetric entropy allocations~\cite[Eq.~(5.1)]{Pip03} and appears as a facet of the symmetrized quantum entropy cone in~\cite[Eq.~(5.2)]{FHC22}. The proof given here derives it directly from SSA.

Pippenger's setting concerns weakly symmetric states, whose subsystem entropies depend only on the number of parties. For a pure state $|\psi\rangle$, the tensor product of its $n!$ party-permuted copies is such a state: every $I_k\in\mathfrak{I}_k$ has entropy $k!\,(n-k)!\, \mathcal{E}_{\mathfrak{I}_k}=n!\,\overline{\mathcal{E}}_k$, since each element of $\mathfrak{I}_k$ occurs $k!\,(n-k)!$ times among the permuted copies. Hence $(\overline{\mathcal{E}}_k)_{k=0}^{n}$ is, up to the factor $n!$, a symmetric allocation of entropy in the sense of~\cite{Pip03}. The proof above avoids this detour by applying SSA directly to $|\psi\rangle$. Together with the endpoint and symmetry conditions, the concavity inequalities in Eq.~\eqref{average-marginal-concavity} describe the closure of the set of average marginal profiles~\cite[Sec.~5]{FHC22}, building on~\cite{Pip03}.
\end{remark}

We now compare the entanglement of an individual cut with the total at its marginal order. Combining the marginal hierarchy in Lemma~\ref{lem:average-marginal-hierarchy} with the MEI in Theorem~\ref{Theorem:MEI} yields an extension of the EPI to arbitrary $k$-marginal entanglement.

\begin{corollary}[$k$-Marginal Entanglement Inequality]\label{cor:k-MEI}
Let $|\psi\rangle\in\medotimes_{j=1}^{n}\mathbbm{C}^{d_j}$ be a pure $n$-partite quantum state, and let $1\leq k\leq\lfloor n/2\rfloor$. Then, for every $I_k\in\mathfrak{I}_k$,
\begin{equation}\label{k-MEI}
\frac{2}{n}\binom{n}{k}\mathcal{E}_{I_k}\leq \mathcal{E}_{\mathfrak{I}_k}.
\end{equation}
\end{corollary}
\begin{proof}
By Lemma~\ref{lem:average-marginal-hierarchy}, $\frac{1}{n}\binom{n}{k}\mathcal{E}_{\mathfrak{I}_1}\leq \mathcal{E}_{\mathfrak{I}_k}$, for every $1\leq k\leq\lfloor n/2\rfloor$. On the other hand, the MEI in Theorem~\ref{Theorem:MEI} gives $2\,\mathcal{E}_{I_k}\leq\mathcal{E}_{\mathfrak{I}_1}$, for every $I_k\in\mathfrak{I}_k$. Combining the two inequalities yields Eq.~\eqref{k-MEI}.
\end{proof}

The preceding corollary follows directly from the marginal hierarchy and the MEI. The next theorem gives a sharper coefficient that is obtained by averaging SSA over subsets on both sides of the distinguished bipartition, followed by pure-state complementarity and binomial counting.

\begin{theorem}[Sharp $k$-Marginal Entanglement Inequality]\label{Theorem:SHMEI}
Let $|\psi\rangle\in\medotimes_{j=1}^{n}\mathbbm{C}^{d_j}$ be a pure $n$-partite quantum state with $n\geq2$, and let $1\leq k\leq n-1$. Then, for every $I_k\in\mathfrak{I}_k$,
\begin{equation}\label{SHMEI}
2\binom{n-2}{k-1}\mathcal{E}_{I_k}\leq
\mathcal{E}_{\mathfrak{I}_k}.
\end{equation}
The coefficient is optimal uniformly over pure states and local dimensions.
\end{theorem}
\begin{proof}
Fix $I_k\in\mathfrak{I}_k$. We use the convention $\binom ab=0$ when $b<0$ or $b>a$. Let $r+s=k$ with $0\leq r\leq k$ and $0\leq s\leq n-k$. For a fixed $K_s\subseteq\bar{I}_k$ of cardinality $s$, averaging SSA over subsets of $I_k$, as in the proof of Lemma~\ref{lem:average-marginal-hierarchy}, gives
\begin{equation}\label{sharp-fixed-complement}
\binom{k-1}{r-1}\mathcal{E}_{I_k\cup K_s}+\binom{k-1}{r}\mathcal{E}_{K_s}
\leq\sum_{\substack{J_r\subseteq I_k\\|J_r|=r}}
\mathcal{E}_{J_r\cup K_s}.
\end{equation}
Indeed, the normalized averages over $J_r$ form a concave sequence with endpoint values $\mathcal{E}_{K_s}$ and $\mathcal{E}_{I_k\cup K_s}$. Each average is therefore bounded below by the corresponding linear interpolation of these endpoints; multiplication by $\binom kr$ yields Eq.~\eqref{sharp-fixed-complement}. Applying the same argument to subsets of $\bar{I}_k$ gives
\begin{align*}
\binom{n-k-1}{s-1}\mathcal{E}_{I_k} &\leq
\sum_{\substack{K_s\subseteq\bar{I}_k\\|K_s|=s}}
\mathcal{E}_{K_s},\\
\binom{n-k-1}{s}\mathcal{E}_{I_k} &\leq
\sum_{\substack{K_s\subseteq\bar{I}_k\\|K_s|=s}}
\mathcal{E}_{I_k\cup K_s},
\end{align*}
where we used $\mathcal{E}_{\emptyset}=\mathcal{E}_{[n]}=0$ and $\mathcal{E}_{\bar{I}_k}=\mathcal{E}_{I_k}$. Summing Eq.~\eqref{sharp-fixed-complement} over $K_s$ and using these two bounds yields
\begin{equation*}
\left[\binom{k-1}{r-1}\binom{n-k-1}{s}+\binom{k-1}{r}\binom{n-k-1}{s-1}
\right]\mathcal{E}_{I_k} \leq
\sum_{\substack{J_r\subseteq I_k\\|J_r|=r}}
\sum_{\substack{K_s\subseteq\bar{I}_k\\|K_s|=s}}\mathcal{E}_{J_r\cup K_s}.
\end{equation*}
Every $J_k\in\mathfrak{I}_k$ has the unique decomposition $J_k=(J_k\cap I_k)\cup(J_k\cap\bar{I}_k)$. Summing over the admissible pairs $r+s=k$ therefore counts every $k$-marginal term exactly once. Moreover, Vandermonde's identity gives
\begin{equation*}
\sum_{r+s=k}\binom{k-1}{r-1}\binom{n-k-1}{s}=
\sum_{r+s=k}\binom{k-1}{r}\binom{n-k-1}{s-1}=\binom{n-2}{k-1}.
\end{equation*}
These identities give Eq.~\eqref{SHMEI}. To show optimality, let one party in $I_k$ and one in $\bar{I}_k$ share a pure entangled state, and put each remaining party in the state $|0\rangle$. A $k$-party marginal has entropy $\mathcal{E}_{I_k}>0$ if it contains exactly one member of the pair, and zero otherwise. Choosing one member of the pair and $k-1$ of the $n-2$ parties outside the pair gives $2\binom{n-2}{k-1}$ marginals with positive entropy. Hence equality holds.
\end{proof}

At the level of set functions, Eq.~\eqref{SHMEI} also follows from a general inequality for nonnegative symmetric submodular functions \cite[Theorem~7(b)]{Filmus13}. That work obtains the coefficient by an uncrossing and term-rewriting method; the proof above instead derives it directly by averaging SSA for quantum marginal entropies.

For $1\leq k\leq\lfloor n/2\rfloor$, the coefficient in Eq.~\eqref{SHMEI} improves that of Eq.~\eqref{k-MEI} by the factor $k(n-k)/(n-1)$. The two coefficients agree for $k=1$, whereas the improvement is strict for $2\leq k\leq\lfloor n/2\rfloor$. At $k=2$, the sharp coefficient is $2(n-2)$, compared with $n-1$ in Corollary~\ref{cor:k-MEI}.

\begin{corollary}[Generalized Entanglement Polygon Inequality]
\label{Corollary:GEPI}
For every pure $n$-partite state, every $1\leq k\leq n-1$, and every
$I_k\in\mathfrak{I}_k$,
\begin{equation}\label{GEPI}
2\,\mathcal{E}_{I_k}\leq\mathcal{E}_{\mathfrak{I}_k}.
\end{equation}
\end{corollary}
\begin{proof}
For $1\leq k\leq\lfloor n/2\rfloor$, one has
$\frac2n\binom nk\geq2$, so Eq.~\eqref{k-MEI} implies the claim.
For larger $k$, apply the same argument at order $n-k$ and
use complementarity.
\end{proof}

For $k=1$ this is the usual EPI, and for $k=2$ it contains the
two-marginal polygon inequality. Thus, the MEI and the average marginal
hierarchy already establish the polygon structure at every order;
Theorem~\ref{Theorem:SHMEI} gives an additional quantitative refinement.

\begin{proposition}\label{prop:GEPI-alpha}
Let $|\psi\rangle\in\medotimes_{j=1}^{n}\mathbbm{C}^{d_j}$ be a pure $n$-partite state with $n\geq2$, and let $\alpha\in(0,1]$. For $n\geq3$, adopt the cyclic convention
$A_{j+n}\equiv A_j$. Then, for every $l\in[n]$,
\begin{equation}\label{EEI-alpha}
2\,\mathcal{E}_{A_lA_{l+1}}^\alpha
\leq\sum_{j=1}^{n}\mathcal{E}_{A_jA_{j+1}}^\alpha.
\end{equation}
For every $1\leq k\leq n-1$ and $I_k\in\mathfrak{I}_k$,
\begin{equation}\label{kMEI-alpha}
2\binom{n-2}{k-1}\mathcal{E}_{I_k}^\alpha
\leq\sum_{J_k\in\mathfrak{I}_k}\mathcal{E}_{J_k}^\alpha.
\end{equation}
In particular,
\begin{equation}\label{GEPI-alpha}
2\,\mathcal{E}_{I_k}^\alpha
\leq\sum_{J_k\in\mathfrak{I}_k}\mathcal{E}_{J_k}^\alpha.
\end{equation}
The coefficient in Eq.~\eqref{kMEI-alpha} is optimal for every
$0<\alpha\leq1$.
\end{proposition}
\begin{proof}
Let $x_1,\ldots,x_N$ be nonnegative quantities satisfying $c\,x_t\leq\sum_{j=1}^N x_j$ for every $t\in[N]$, with a common coefficient $c>0$. Set $x_{\mathrm{max}}\coloneqq\max\{x_j\mid j\in[N]\}$. If $x_{\mathrm{max}}=0$, the powered inequalities are immediate. Otherwise, $0\leq x_j/x_{\mathrm{max}}\leq1$, and $u\leq u^\alpha$ for $u\in[0,1]$. Applying the hypothesis at an index attaining $x_{\mathrm{max}}$ gives, for every $t\in[N]$,
\begin{equation*}
c\,x_t^\alpha \leq c\,x_{\mathrm{max}}^\alpha \leq
x_{\mathrm{max}}^{\alpha-1}\sum_{j=1}^{N}x_j=x_{\mathrm{max}}^\alpha\sum_{j=1}^{N}\frac{x_j}{x_{\mathrm{max}}}\leq
x_{\mathrm{max}}^\alpha\sum_{j=1}^{N}\Bigl(\frac{x_j}{x_{\mathrm{max}}}\Bigr)^\alpha=\sum_{j=1}^{N}x_j^\alpha.
\end{equation*}
Apply this observation to the EEI, with $N=n$, the quantities $\mathcal{E}_{A_jA_{j+1}}$, and $c=2$, and to Theorem~\ref{Theorem:SHMEI}, with $N=\binom{n}{k}$, the quantities $\mathcal{E}_{J_k}$ for $J_k\in\mathfrak{I}_k$, and $c=2\binom{n-2}{k-1}$. This proves Eqs.~\eqref{EEI-alpha} and~\eqref{kMEI-alpha}. Since $2\binom{n-2}{k-1}\geq2$, Eq.~\eqref{GEPI-alpha} follows. For the state used to prove optimality in Theorem~\ref{Theorem:SHMEI}, every nonzero $k$-marginal entanglement equals $\mathcal{E}_{I_k}>0$. Hence the same state saturates Eq.~\eqref{kMEI-alpha} for every $\alpha\in(0,1]$.
\end{proof}

\subsection{Graph and Hypergraph Inequalities}
\label{subsec:graph-hypergraph}

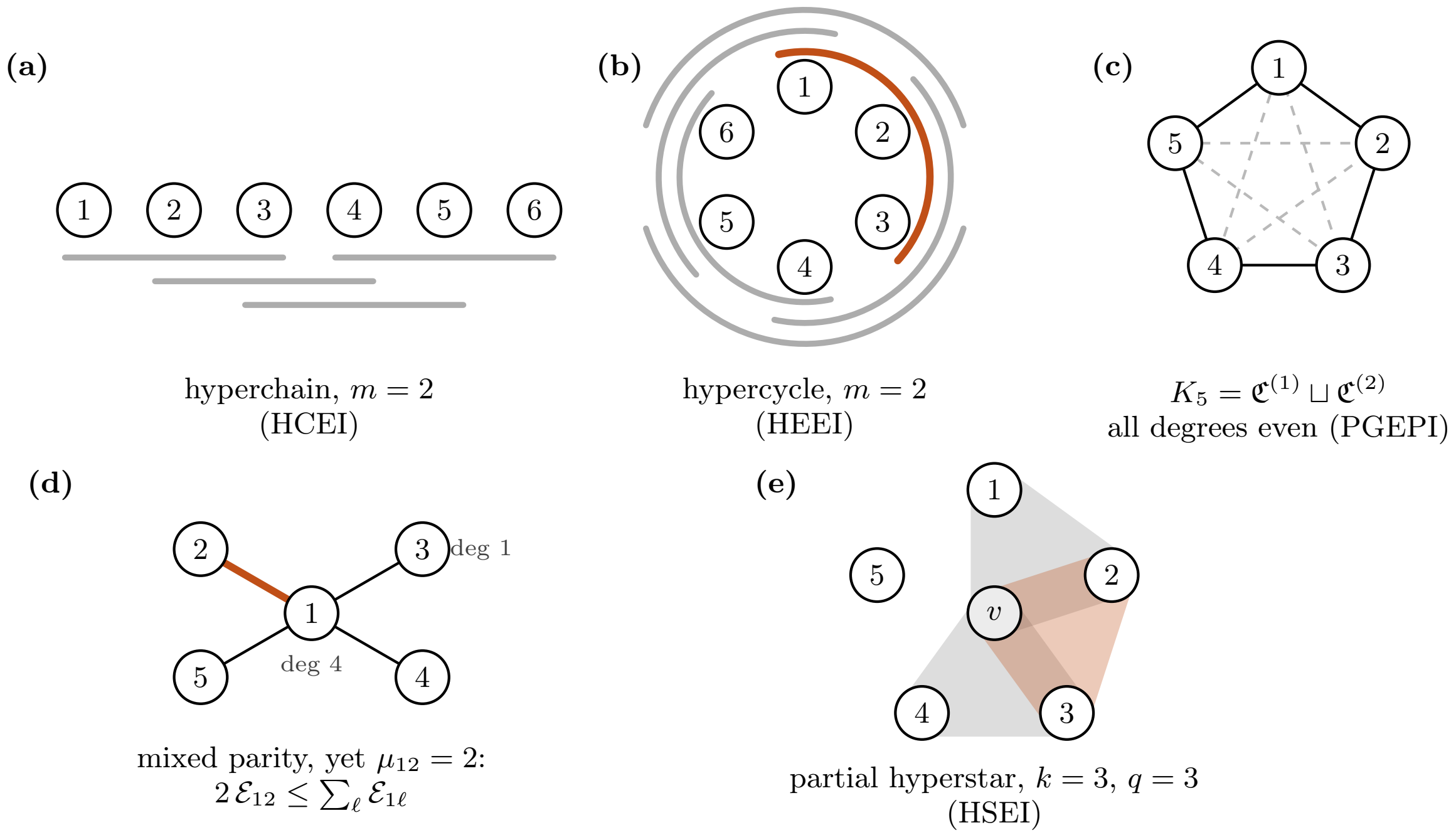
\begin{figure*}[t]
\centering
\begin{tikzpicture}[
  v/.style={circle,draw,thick,fill=white,minimum size=5.6mm,inner sep=0pt,font=\small},
  plabel/.style={font=\bfseries},
  caption text/.style={font=\small,align=center}
]

\begin{scope}
  \node[plabel] at (-0.6,1.15) {(a)};
  \foreach \x in {1,...,6} \coordinate (a\x) at ({(\x-1)*0.95},-0.35);
  \foreach \jw/\y in {1/-0.85,2/-1.1,3/-1.35,4/-0.85} {
    \pgfmathtruncatemacro{\idx}{\jw+2}
    \draw[line width=1.8pt,line cap=round,gray!65]
      ($(a\jw)+(-0.2,0)$) ++(0,\y+0.35) -- ($(a\idx)+(0.2,\y+0.35)$);
  }
  \foreach \x in {1,...,6} \node[v] at (a\x) {$\x$};
  \node[caption text] at (2.375,-2.45) {hyperchain, $m=2$\\(HCEI)};
\end{scope}

\begin{scope}[xshift=7.6cm]
  \node[plabel] at (-1.95,1.15) {(b)};
  \foreach \kk in {1,...,6} \coordinate (b\kk) at ({90-(\kk-1)*60}:0.95);
  \foreach \kk/\rad in {1/1.32,2/1.54,3/1.76,4/1.32,5/1.54,6/1.76} {
    \pgfmathsetmacro{\angA}{90-(\kk-1)*60+12}
    \pgfmathsetmacro{\angB}{90-(\kk+1)*60-12}
    \draw[line width=1.8pt,line cap=round,gray!65]
      (\angA:\rad) arc[start angle=\angA,end angle=\angB,radius=\rad];
  }
  \pgfmathsetmacro{\angA}{90+12}
  \pgfmathsetmacro{\angB}{90-120-12}
  \draw[line width=2.2pt,line cap=round,figaccent]
    (\angA:1.32) arc[start angle=\angA,end angle=\angB,radius=1.32];
  \foreach \kk in {1,...,6} \node[v] at (b\kk) {$\kk$};
  \node[caption text] at (0,-2.45) {hypercycle, $m=2$\\(HEEI)};
\end{scope}

\begin{scope}[xshift=12.6cm]
  \node[plabel] at (-1.75,1.15) {(c)};
  \foreach \kk in {1,...,5} \coordinate (c\kk) at ({90-(\kk-1)*72}:1.15);
  \draw[thick] (c1)--(c2)--(c3)--(c4)--(c5)--cycle;
  \draw[thick,gray!55,dashed] (c1)--(c3)--(c5)--(c2)--(c4)--cycle;
  \foreach \kk in {1,...,5} \node[v] at (c\kk) {$\kk$};
  \node[caption text] at (0,-2.45) {$K_5=\mathfrak{C}^{(1)}\sqcup\mathfrak{C}^{(2)}$\\all degrees even (PGEPI)};
\end{scope}

\begin{scope}[xshift=2.4cm,yshift=-4.6cm]
  \node[plabel] at (-2.75,1.35) {(d)};
  \coordinate (d1) at (0,0);
  \foreach \kk/\dang in {2/150,3/30,4/-30,5/-150} \coordinate (d\kk) at (\dang:1.35);
  \foreach \kk in {3,4,5} \draw[thick] (d1)--(d\kk);
  \draw[figaccent,line width=2.2pt] (d1)--(d2);
  \foreach \kk in {1,...,5} \node[v] at (d\kk) {$\kk$};
  \node[font=\scriptsize,gray!50!black] at (0.0,-0.55) {deg $4$};
  \node[font=\scriptsize,gray!50!black] at ($(d3)+(0.62,0)$) {deg $1$};
  \node[caption text] at (0,-1.75) {mixed parity, yet $\mu_{12}=2$:\\$2\,\mathcal{E}_{12}\leq\sum_{\ell}\mathcal{E}_{1\ell}$};
\end{scope}

\begin{scope}[xshift=9.6cm,yshift=-4.6cm]
  \node[plabel] at (-2.3,1.35) {(e)};
  \coordinate (ea) at (0,0);
  \foreach \kk/\ang in {1/90,2/18,3/-54,4/-126,5/162} \coordinate (e\kk) at (\ang:1.3);
  \begin{scope}[transparency group,opacity=0.22]
    \filldraw[gray!80!black,line width=5mm,line join=round] (ea)--(e1)--(e2)--cycle;
  \end{scope}
  \begin{scope}[transparency group,opacity=0.22]
    \filldraw[gray!80!black,line width=5mm,line join=round] (ea)--(e3)--(e4)--cycle;
  \end{scope}
  \begin{scope}[transparency group,opacity=0.30]
    \filldraw[figaccent,line width=5mm,line join=round] (ea)--(e2)--(e3)--cycle;
  \end{scope}
  \node[v,fill=gray!15] at (ea) {$v$};
  \foreach \kk in {1,...,5} \node[v] at (e\kk) {$\kk$};
  \node[caption text] at (0,-1.95) {partial hyperstar, $k=3$, $q=3$\\(HSEI)};
\end{scope}

\end{tikzpicture}
\caption{Graph and hypergraph families of cuts. Vertices are parties, and each edge or hyperedge selects the cut between its parties and the rest. (a)~Hyperchain with $m=2$: consecutive windows $\{j,j+1,j+2\}$ overlap in two parties (HCEI). (b)~Hypercycle with $m=2$ on six parties: every party lies in three windows, and the highlighted window is the distinguished cut of the HEEI. (c)~Walecki decomposition of $K_5$ into two Hamiltonian cycles (solid and dashed); all vertex degrees are even, so the PGEPI applies. (d)~Five-party star with distinguished edge $\{1,2\}$: the hub has even degree and the leaves odd degree, so the parity condition fails, yet $\mu_{12}=2$ in the WTPCEI (Theorem~\ref{theorem:weighted-two-party-cut}), and the polygon inequality with coefficient two holds. (e)~Partial hyperstar with $k=3$ and $q=3$ centered at $v$; hyperedges may overlap outside the center (HSEI).}
\label{fig:hypergraph-families}
\end{figure*}

The CEI and the EEI associate marginal entanglement constraints with, respectively, paths and cycles of subsystem labels. This viewpoint suggests replacing ordinary edges by larger overlapping subsets and, more generally, representing marginal entanglement by hyperedges of a hypergraph (Fig.~\ref{fig:hypergraph-families}). We first extend the chain construction to consecutive higher-order marginals and then develop a graph-theoretic formulation that applies to considerably more general edge and hyperedge families.

We begin with the higher-order analogue of the CEI. Consecutive $(m+1)$-party marginals now play the role of edges, while their $m$-party overlaps play the role of the intermediate vertices of the ordinary chain.

\begin{theorem}[Hyperchain Entanglement Inequality]
\label{Theorem:HCEI}
Let $|\psi\rangle\in\medotimes_{j=1}^{n}\mathbbm{C}^{d_j}$ be a pure $n$-partite quantum state, and let $1\leq m\leq n-2$. Then
\begin{equation}\label{HCEI-deficit}
\sum_{j=1}^{n-m}\mathcal{E}_{A_j\cdots A_{j+m}} -
\sum_{j=1}^{n-m-1}\mathcal{E}_{A_{j+1}\cdots A_{j+m}} =
\sum_{j=1}^{n-m-1}\mathcal{I}\big(A_1\cdots A_j:A_{j+m+1}
\mid A_{j+1}\cdots A_{j+m}\big).
\end{equation}
Consequently,
\begin{equation}\label{HCEI}
\sum_{j=1}^{n-m-1}
\mathcal{E}_{A_{j+1}\cdots A_{j+m}} \leq
\sum_{j=1}^{n-m}\mathcal{E}_{A_j\cdots A_{j+m}}.
\end{equation}
Moreover, equality holds if and only if
\begin{equation}\label{Markov-chain-HC}
\mathcal{I}\big(A_1\cdots A_j:A_{j+m+1}\mid A_{j+1}\cdots A_{j+m}\big)=0,
\qquad \forall~1\leq j\leq n-m-1.
\end{equation}
\end{theorem}
\begin{proof}
For every $1\leq j\leq n-m-1$, the definition of conditional mutual
information gives
\begin{align*}
\mathcal{I}\big(A_1\cdots A_j:A_{j+m+1}\mid A_{j+1}\cdots A_{j+m}\big)
&=\mathcal{S}(\varrho_{A_1\cdots A_{j+m}})
+\mathcal{S}(\varrho_{A_{j+1}\cdots A_{j+m+1}})\\
&\quad-\mathcal{S}(\varrho_{A_{j+1}\cdots A_{j+m}})
-\mathcal{S}(\varrho_{A_1\cdots A_{j+m+1}}).
\end{align*}
Summing over $j=1,\ldots,n-m-1$, the prefix entropies telescope:
\begin{equation*}
\sum_{j=1}^{n-m-1}\mathcal{I}\big(A_1\cdots A_j:A_{j+m+1}\mid A_{j+1}\cdots A_{j+m}\big) =
\sum_{j=1}^{n-m}\mathcal{S}(\varrho_{A_j\cdots A_{j+m}}) -
\sum_{j=1}^{n-m-1}\mathcal{S}(\varrho_{A_{j+1}\cdots A_{j+m}}) -
\mathcal{S}(\varrho_{A_1\cdots A_n}).
\end{equation*}
Since the global state is pure, $\mathcal{S}(\varrho_{A_1\cdots A_n})=0$, and the remaining subsystem entropies are precisely the corresponding marginal entanglement. This proves Eq.~\eqref{HCEI-deficit}. By SSA, every conditional mutual information on the right-hand side is nonnegative, and hence Eq.~\eqref{HCEI} follows. Equality holds if and only if every term on the right-hand side of Eq.~\eqref{HCEI-deficit} vanishes, which gives Eq.~\eqref{Markov-chain-HC}.
\end{proof}

We denote the common quantity in Eq.~\eqref{HCEI-deficit} by $\Delta_m$. In particular, $\Delta_1=\Delta_{\mathrm{CEI}}$.

\begin{remark}
For $m=1$, Theorem~\ref{Theorem:HCEI} reduces exactly to the CEI in Theorem~\ref{Theorem:CEI}. Thus, the HCEI extends the path inequality from ordinary edges to consecutive hyperedges with increasingly large overlaps.
\end{remark}

The cyclic analogue of this construction is most naturally formulated in terms of the symmetric difference introduced in Subsec.~\ref{subsec:graph}. This operation provides a natural way of encoding the parity of subsystem incidence and leads to polygon inequalities for general collections of overlapping hyperedges, rather than only for consecutive ones.

\begin{lemma}[Symmetric-Difference Entropy Triangle Inequality]\label{lem:symmetric-difference}
For an arbitrary multipartite density operator and $I,J\subseteq[n]$,
\begin{equation}\label{symmetric-difference-triangle}
\mathcal{S}(\varrho_{I\triangle J}) \leq
\mathcal{S}(\varrho_I)+\mathcal{S}(\varrho_J).
\end{equation}
For pure states, this is the corresponding inequality for marginal entanglement.
\end{lemma}
\begin{proof}
Set $X=I\setminus J$, $Y=J\setminus I$, and $Z=I\cap J$.
SA followed by WM gives
\begin{equation*}
\mathcal{S}(\varrho_{XY})
\leq\mathcal{S}(\varrho_X)+\mathcal{S}(\varrho_Y)
\leq\mathcal{S}(\varrho_{XZ})+\mathcal{S}(\varrho_{YZ}).
\end{equation*}
Empty subsystems have entropy zero. No global purity is needed.
\end{proof}

Repeated application of Lemma~\ref{lem:symmetric-difference} produces polygon inequalities for families of subsystem sets satisfying a simple parity condition.

\begin{theorem}[Parity-Hypergraph Entanglement Polygon Inequality]\label{Theorem:parity-hypergraph}
Let $|\psi\rangle\in\medotimes_{j=1}^{n}\mathbbm{C}^{d_j}$ be a pure $n$-partite quantum state. Let $\{I_{k_1}^{(1)},\ldots,I_{k_q}^{(q)}\}$ be a family of nonempty proper subsets of $[n]$, where $I_{k_t}^{(t)}\in\mathfrak{I}_{k_t}$ and $1\leq k_t\leq n-1$ for every $t\in[q]$. Suppose that
\begin{equation}\label{parity-hypergraph-condition}
I_{k_1}^{(1)}\triangle\cdots\triangle I_{k_q}^{(q)}\in\{\emptyset,[n]\}.
\end{equation}
Then, for every $s\in[q]$,
\begin{equation}\label{parity-hypergraph-EPI}
2\,\mathcal{E}_{I_{k_s}^{(s)}}\leq\sum_{t=1}^{q}\mathcal{E}_{I_{k_t}^{(t)}}.
\end{equation}
\end{theorem}
\begin{proof}
Fix $s\in[q]$. If $\bigtriangleup_{t=1}^{q}I_{k_t}^{(t)}=\emptyset$, then $\bigtriangleup_{\substack{1\leq t\leq q\\ t\neq s}}\,I_{k_t}^{(t)}=I_{k_s}^{(s)}$. If instead $\bigtriangleup_{t=1}^{q}I_{k_t}^{(t)}=[n]$, then $\bigtriangleup_{\substack{1\leq t\leq q\\ t\neq s}}\,I_{k_t}^{(t)}=\bar{I}_{k_s}^{(s)}$. Repeated application of Lemma~\ref{lem:symmetric-difference}, together with complementarity in the second case, therefore gives $\mathcal{E}_{I_{k_s}^{(s)}} \leq\sum_{\substack{1\leq t\leq q\\ t\neq s}}\mathcal{E}_{I_{k_t}^{(t)}}$. Adding $\mathcal{E}_{I_{k_s}^{(s)}}$ to both sides proves Eq.~\eqref{parity-hypergraph-EPI}.
\end{proof}

\begin{remark}
When the total symmetric difference is empty, the same proof yields a
polygon inequality for marginal entropies even for a mixed global state.
When the symmetric difference is $[n]$, the entanglement version uses
pure-state complementarity. In both cases, the parity condition is
sufficient; no necessity claim is made.
\end{remark}

The condition in
Eq.~\eqref{parity-hypergraph-condition} has a particularly simple
interpretation for ordinary graphs: the symmetric difference of all
edges is precisely the set of vertices having odd degree.

\begin{corollary}[Parity-Graph Entanglement Polygon Inequality]\label{cor:parity-graph}
Let $|\psi\rangle$ be a pure $n$-partite state with $n\geq3$, and let $G=([n],E)$ be a graph such that either every vertex has even degree or every vertex has odd degree. Then, for every edge $\{k,l\}\in E$,
\begin{equation}\label{parity-graph-EPI}
2\,\mathcal{E}_{A_kA_l} \leq
\sum_{\{i,j\}\in E}\mathcal{E}_{A_iA_j}.
\end{equation}
\end{corollary}
\begin{proof}
The symmetric difference of all edges of $G$ is the set of vertices having odd degree. It is therefore $\emptyset$ when every degree is even and $[n]$ when every degree is odd. The claim follows immediately from Theorem~\ref{Theorem:parity-hypergraph}.
\end{proof}

\begin{remark}
The EEI is recovered from Corollary~\ref{cor:parity-graph} by taking $G=C_n$, since every vertex of a cycle has degree two. Thus, the PGEPI extends the EEI from a single cycle to every graph whose vertex degrees have uniform parity.
\end{remark}

\begin{remark}\label{rem:TMEPI-parity-graph}
The TMEPI in Corollary~\ref{cor:2-EPI} can also be obtained directly from Corollary~\ref{cor:parity-graph}. Indeed, the complete graph $K_n$ is $(n-1)$-regular, and therefore all of its vertices have the same degree parity. Hence, applying Corollary~\ref{cor:parity-graph} to $K_n$ recovers Eq.~\eqref{2-EPI}. Thus, besides its derivation from the EEI through a Hamiltonian cycle, the TMEPI also arises as the complete-graph specialization of the PGEPI.
\end{remark}

The same parity principle also applies directly to uniform hypergraphs. In particular, closing a hyperchain into a cyclic family of consecutive hyperedges yields the following higher-order analogue of the EEI.

\begin{corollary}[Hyperedge Entanglement Inequality]
\label{Theorem:HEEI}
Let $|\psi\rangle\in\medotimes_{j=1}^{n}\mathbbm{C}^{d_j}$ be a pure $n$-partite quantum state. Fix $1\leq m\leq n-2$ and adopt the cyclic convention for the subsystem labels. Then, for every $k\in[n]$,
\begin{equation}\label{HEEI}
2\,\mathcal{E}_{A_kA_{k+1}\cdots A_{k+m}}\leq
\sum_{j=1}^{n}\mathcal{E}_{A_jA_{j+1}\cdots A_{j+m}}.
\end{equation}
The coefficient two is optimal for every admissible $n,m$, uniformly over pure states and local dimensions.
\end{corollary}
\begin{proof}
For each $j\in[n]$, define $(m+1)$-subsets $I_{m+1}^{(j)}\coloneqq\{j,j+1,\ldots,j+m\}\in\mathfrak{I}_{m+1}$, with indices interpreted cyclically modulo $n$. These are $n$ distinct subsets because $m+1<n$. Each vertex $v\in[n]$ belongs precisely to the windows $I_{m+1}^{(v)},I_{m+1}^{(v-1)},\ldots,I_{m+1}^{(v-m)}$, and hence occurs in exactly $m+1$ members of the family. Since the symmetric difference contains precisely the vertices occurring an odd number of times,
\begin{equation*}
\bigtriangleup_{j=1}^{n}I_{m+1}^{(j)}=\begin{cases}
\emptyset, & m+1 \text{ is even},\\
[n], & m+1 \text{ is odd}.
\end{cases}
\end{equation*}
The family therefore satisfies Eq.~\eqref{parity-hypergraph-condition}, and the PHEPI in Theorem~\ref{Theorem:parity-hypergraph} gives Eq.~\eqref{HEEI}. To prove optimality, fix the distinguished window $A_kA_{k+1}\cdots A_{k+m}$ and place a pure entangled pair of entropy $h>0$ on $A_{k+m}A_{k+m+1}$. Put all remaining parties in pure states. Exactly two cyclic windows contain precisely one endpoint of the pair and therefore have entropy $h$; all other windows have entropy zero. Consequently,
\begin{equation*}
\sum_{j=1}^{n}
\mathcal{E}_{A_jA_{j+1}\cdots A_{j+m}}=2h
=2\,\mathcal{E}_{A_kA_{k+1}\cdots A_{k+m}}.
\end{equation*}
Thus equality is attained with nonzero distinguished entanglement, and the coefficient cannot be increased.
\end{proof}

\begin{remark}
For $m=1$, the HEEI reduces exactly to the EEI. Thus, the HCEI and HEEI extend the path--cycle pair formed by the CEI and EEI to overlapping higher-order hyperchains and hypercycles.
\end{remark}

The polygon inequalities above also admit powered forms, with the same coefficient, for every $\alpha\in(0,1]$.
\begin{proposition}\label{prop:parity-HEEI-alpha}
Let $|\psi\rangle\in\medotimes_{j=1}^{n}\mathbbm{C}^{d_j}$ be a pure $n$-partite state with $n\geq2$, and let $\alpha\in(0,1]$. For any family $\{I_{k_1}^{(1)},\ldots,I_{k_q}^{(q)}\}$ satisfying the assumptions of Theorem~\ref{Theorem:parity-hypergraph} and every $s\in[q]$,
\begin{equation}\label{parity-hypergraph-EPI-alpha}
2\,\mathcal{E}_{I_{k_s}^{(s)}}^\alpha \leq
\sum_{t=1}^{q}\mathcal{E}_{I_{k_t}^{(t)}}^\alpha.
\end{equation}
In particular, for $n\geq3$, adopt the cyclic convention
$A_{j+n}\equiv A_j$. Then, for every
$1\leq m\leq n-2$ and $k\in[n]$,
\begin{equation}\label{HEEI-alpha}
2\,\mathcal{E}_{A_kA_{k+1}\cdots A_{k+m}}^\alpha \leq
\sum_{j=1}^{n}\mathcal{E}_{A_jA_{j+1}\cdots A_{j+m}}^\alpha.
\end{equation}
\end{proposition}
\begin{proof}
Fix $s\in[q]$. Subtracting the distinguished term from Eq.~\eqref{parity-hypergraph-EPI} gives
\begin{equation*}
\mathcal{E}_{I_{k_s}^{(s)}} \leq
\sum_{\substack{1\leq t\leq q\\t\neq s}}
\mathcal{E}_{I_{k_t}^{(t)}}.
\end{equation*}
For $0<\alpha\leq1$, the function $x\mapsto x^\alpha$
is increasing and subadditive on $[0,\infty)$. Hence
\begin{equation*}
\mathcal{E}_{I_{k_s}^{(s)}}^\alpha\leq
\Bigl(\sum_{\substack{1\leq t\leq q\\t\neq s}}
\mathcal{E}_{I_{k_t}^{(t)}}\Bigr)^{\!\alpha}\leq
\sum_{\substack{1\leq t\leq q\\t\neq s}}
\mathcal{E}_{I_{k_t}^{(t)}}^\alpha.
\end{equation*}
Adding $\mathcal{E}_{I_{k_s}^{(s)}}^\alpha$ to both sides proves Eq.~\eqref{parity-hypergraph-EPI-alpha}. The cyclic windows satisfy the same parity condition, as shown in the proof of Corollary~\ref{Theorem:HEEI}, so Eq.~\eqref{HEEI-alpha} follows. The equality construction in Corollary~\ref{Theorem:HEEI} has exactly two nonzero window entropies, both equal to $h$. It therefore also saturates Eq.~\eqref{HEEI-alpha} for every $\alpha\in(0,1]$, so the coefficient two remains optimal.
\end{proof}

For $m=1$, Eq.~\eqref{HEEI-alpha} recovers the powered
EEI in Eq.~\eqref{EEI-alpha}.

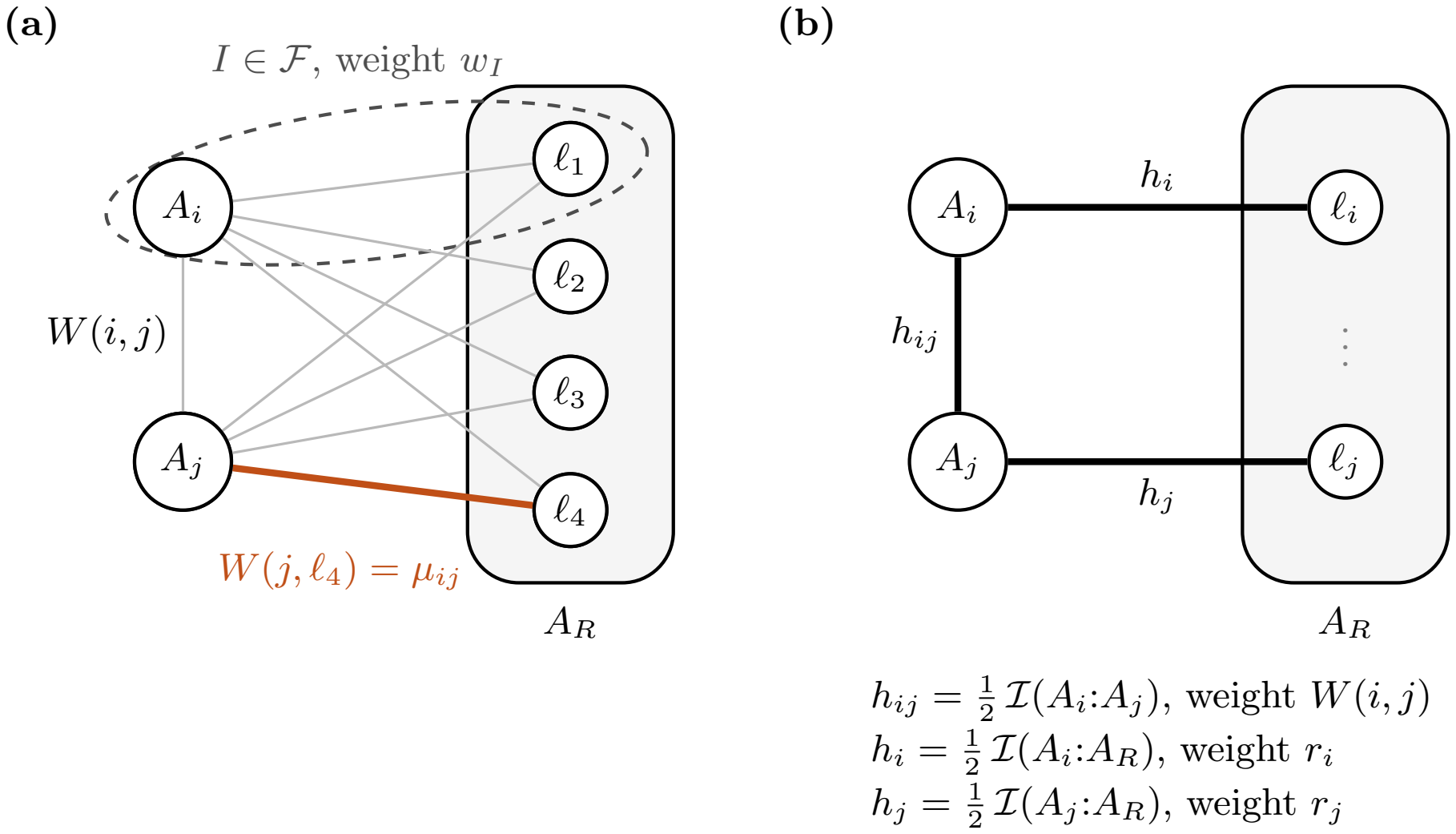
\begin{figure*}[t]
\centering
\begin{tikzpicture}[
  party/.style={circle,draw,thick,fill=white,minimum size=8mm,inner sep=0pt},
  rparty/.style={circle,draw,thick,fill=white,minimum size=6mm,inner sep=0pt,font=\small},
  sep/.style={gray!55,line width=0.6pt},
  bond/.style={line width=1.6pt},
  plabel/.style={font=\bfseries}
]

\begin{scope}
  \node[plabel] at (-1.25,2.55) {(a)};

  \draw[thick,fill=gray!8,rounded corners=12pt] (2.35,-2.05) rectangle (4.05,2.05);
  \node at (3.2,-2.38) {$A_R$};

  \node[party] (i) at (0,1.05) {$A_i$};
  \node[party] (j) at (0,-1.05) {$A_j$};
  \node[rparty] (l1) at (3.2,1.45) {$\ell_1$};
  \node[rparty] (l2) at (3.2,0.48) {$\ell_2$};
  \node[rparty] (l3) at (3.2,-0.48) {$\ell_3$};
  \node[rparty] (l4) at (3.2,-1.45) {$\ell_4$};

  \draw[thick,dashed,gray!60!black,rotate around={7.1:(1.6,1.25)}]
    (1.6,1.25) ellipse (2.25cm and 0.62cm);
  \node[gray!60!black,font=\small] at (1.45,2.25) {$I\in\mathcal{F}$, weight $w_I$};

  \foreach \lab in {l1,l2,l3,l4} {
    \draw[sep] (i) -- (\lab);
    \draw[sep] (j) -- (\lab);
  }
  \draw[sep] (i) -- node[left,black,font=\small] {$W(i,j)$} (j);

  \draw[figaccent,line width=1.6pt] (j) -- (l4);
  \node[figaccent,font=\small,align=center] at (1.3,-1.95) {$W(j,\ell_4)=\mu_{ij}$};

  \node[party] at (i) {$A_i$};
  \node[party] at (j) {$A_j$};
  \foreach \lab/\tx in {l1/\ell_1,l2/\ell_2,l3/\ell_3,l4/\ell_4}
    \node[rparty] at (\lab) {$\tx$};
\end{scope}

\begin{scope}[xshift=6.4cm]
  \node[plabel] at (-1.25,2.55) {(b)};

  \draw[thick,fill=gray!8,rounded corners=12pt] (2.35,-2.05) rectangle (4.05,2.05);
  \node at (3.2,-2.38) {$A_R$};

  \node[party] (i) at (0,1.05) {$A_i$};
  \node[party] (j) at (0,-1.05) {$A_j$};
  \node[rparty] (li) at (3.2,1.05) {$\ell_i$};
  \node[rparty] (lj) at (3.2,-1.05) {$\ell_j$};
  \node[font=\small,gray] at (3.2,0) {$\vdots$};

  \draw[bond] (i) -- node[left,font=\small] {$h_{ij}$} (j);
  \draw[bond] (i) -- node[above,font=\small] {$h_{i}$} (li);
  \draw[bond] (j) -- node[below,font=\small] {$h_{j}$} (lj);

  \node[font=\small,align=left,anchor=north] at (1.6,-2.7) {%
    $h_{ij}=\tfrac12\,\mathcal{I}(A_i{:}A_j)$, weight $W(i,j)$\\[1pt]
    $h_{i}=\tfrac12\,\mathcal{I}(A_i{:}A_R)$, weight $r_i$\\[1pt]
    $h_{j}=\tfrac12\,\mathcal{I}(A_j{:}A_R)$, weight $r_j$};
\end{scope}

\end{tikzpicture}
\caption{The WTPCEI in Theorem~\ref{theorem:weighted-two-party-cut}. (a)~Each cut $I\in\mathcal{F}$ with weight $w_I$ (dashed) contributes $w_I$ to the separation weight $W(u,v)$ of every pair of parties that it separates. Gray lines join the distinguished parties $A_i$ and $A_j$ to the parties $\ell\in R$ of the rest $A_R$. The optimal coefficient $\mu_{ij}$ is the smallest separation weight among these pairs (highlighted). (b)~At fixed $(\mathcal{E}_{A_i},\mathcal{E}_{A_j},\mathcal{E}_{A_iA_j})$, three independent pure bonds attain the minimum of the weighted sum. The bond entropies are one half of the mutual information terms of the tripartition $A_i\,|\,A_j\,|\,A_R$, each bond is weighted by the separation weight of its endpoints, and $\ell_i,\ell_j\in R$ attain $r_i$ and $r_j$.}
\label{fig:weighted-cut}
\end{figure*}

The preceding parity criterion certifies polygon inequalities for families of subsystem cuts. For a distinguished two-party cut and an arbitrary weighted family, we can determine the optimal coefficient exactly (see Fig.~\ref{fig:weighted-cut}). It is set by the total weight of cuts separating pairs of parties on opposite sides of the distinguished cut.

\begin{theorem}[Weighted Two-Party Cut Entanglement Inequality]\label{theorem:weighted-two-party-cut}
Let $|\psi\rangle\in\medotimes_{j=1}^{n}\mathbbm{C}^{d_j}$ be a pure $n$-partite quantum state with $n\geq3$. Fix distinct $i,j\in[n]$, and set $R\coloneqq[n]\setminus\{i,j\}$. Let $\mathcal{F}\subseteq\{I\subsetneq[n]\mid I\neq\emptyset\}$ be a family of cuts with weights $w_I\geq0$. For distinct $u,v\in[n]$, let
\begin{equation*}
W(u,v)\coloneqq\sum_{\substack{I\in\mathcal{F}\\ |I\cap\{u,v\}|=1}}w_I
\end{equation*}
be the total weight of the cuts separating $u$ and $v$. Set $r_x\coloneqq\min_{\ell\in R}W(x,\ell)$ for $x\in\{i,j\}$, and $\mu_{ij}\coloneqq\min\{r_i,r_j\}$. Then
\begin{equation}\label{weighted-two-party-cut}
\mu_{ij}\,\mathcal{E}_{A_iA_j}\leq \sum_{I\in\mathcal{F}}w_I\,\mathcal{E}_I.
\end{equation}
The coefficient $\mu_{ij}$ is optimal for the prescribed weighted family, uniformly over pure states and local dimensions.

More precisely, write $A_R$ for the composite subsystem indexed by $R$. Then
\begin{equation}\label{weighted-two-party-refinement}
W(i,j)\,\mathcal{I}(A_i:A_j)+r_i\,\mathcal{I}(A_i:A_R)+r_j\,\mathcal{I}(A_j:A_R) \leq 2\sum_{I\in\mathcal{F}}w_I\,\mathcal{E}_I.
\end{equation}
By complementarity, the left-hand side depends only on the triple $(\mathcal{E}_{A_i},\mathcal{E}_{A_j},\mathcal{E}_{A_iA_j})$. For every realizable triple, one half of it is the minimum of $\sum_{I\in\mathcal{F}}w_I\mathcal{E}_I$ over all pure states with that triple, allowing arbitrary finite local dimensions, and a product of at most three bipartite pure states attains this minimum.
\end{theorem}
\begin{proof}
Fix an order on $R$, and let $P_\ell$ and $Q_\ell$ be the sets preceding and following $\ell$, respectively. For $\ell\notin U$, write $D(\ell\,|\,U)\coloneqq\mathcal{E}_{U\cup\{\ell\}}-\mathcal{E}_U$. For $B\subseteq\{i,j\}$, set $L_\ell(B)\coloneqq[D(\ell\,|\,P_\ell\cup B)+D(\ell\,|\,Q_\ell\cup B)]/2$. For $\ell\in U\subseteq R$, SSA gives $D(\ell\,|\,P_\ell\cup B)\leq D(\ell\,|\,(P_\ell\cap U)\cup B)$, and the same inequality holds with $Q_\ell$ in place of $P_\ell$. The sets $P_\ell\cap U$ and $Q_\ell\cap U$ are the predecessor sets in the forward and reverse orders on $U$, respectively. The two restricted sums therefore telescope, giving $\sum_{\ell\in U}L_\ell(B)\leq\mathcal{E}_{U\cup B}-\mathcal{E}_B$, with equality for $U=R$. By complementarity, $D(\ell\,|\,P_\ell\cup\{i\})=-D(\ell\,|\,Q_\ell\cup\{j\})$, and likewise with $P_\ell$ and $Q_\ell$ exchanged. Hence $L_\ell(\{j\})=-L_\ell(\{i\})$. Assign weights $\lambda_{ij}=\mathcal{I}(A_i:A_j)/2$ and $\lambda_{x\ell}=[L_\ell(\emptyset)-L_\ell(\{x\})]/2$ for $x\in\{i,j\}$ and $\ell\in R$, with all other weights zero. These weights are nonnegative by SSA and the nonnegativity of mutual information. The equality for $U=R$ gives $\sum_{\ell\in R}\lambda_{x\ell}=\mathcal{I}(A_x:A_R)/2$. Let $C_\lambda(I)$ be the total pair weight crossing $I\,|\,\bar{I}$. For $I\subseteq R$, we have $C_\lambda(I)=\sum_{\ell\in I}L_\ell(\emptyset)\leq\mathcal{E}_I$. For $I=\{i\}\cup U$ with $U\subseteq R$, we have
\begin{equation*}
C_\lambda(I)=\lambda_{ij}+\sum_{\ell\in R}\lambda_{i\ell}
+\sum_{\ell\in U}L_\ell(\{i\})
=\mathcal{E}_{A_i}+\sum_{\ell\in U}L_\ell(\{i\})
\leq\mathcal{E}_I.
\end{equation*}
Taking complements covers the remaining cuts. Summing with weights $w_I$ and exchanging the order of summation gives
\begin{equation*}
\sum_{I\in\mathcal{F}}w_I\,\mathcal{E}_I
\geq\sum_{\{u,v\}\in\mathfrak{I}_2}\lambda_{uv}W(u,v).
\end{equation*}
Using $W(x,\ell)\geq r_x$ for $x\in\{i,j\}$ proves Eq.~\eqref{weighted-two-party-refinement}. Since $r_i,r_j\geq\mu_{ij}$ and $\mathcal{I}(A_i:A_R)+\mathcal{I}(A_j:A_R)=2\mathcal{E}_{A_iA_j}$, this also gives Eq.~\eqref{weighted-two-party-cut}. For equality in Eq.~\eqref{weighted-two-party-refinement}, choose $\ell_i,\ell_j\in R$ attaining $r_i,r_j$ and place independent pure bonds on $(i,j)$, $(i,\ell_i)$, and $(j,\ell_j)$. Give them entropies equal to half the three mutual information terms, respectively, using separate local tensor factors for any shared party. Across the tripartition $A_i\,|\,A_j\,|\,A_R$, the two bond entropies incident to each block sum to its prescribed entropy. Each bond contributes its entropy to every cut separating its endpoints, giving equality. Since this state has the prescribed triple, one half of the left-hand side of Eq.~\eqref{weighted-two-party-refinement} is the stated minimum. Finally, a single entangled pair between $x\in\{i,j\}$ and $\ell\in R$ with $W(x,\ell)=\mu_{ij}$ attains Eq.~\eqref{weighted-two-party-cut} with $\mathcal{E}_{A_iA_j}>0$, proving optimality.
\end{proof}

For unit weights and $\{i,j\}\in\mathcal{F}$, optimality of $\mu_{ij}$ shows that the polygon inequality with coefficient two holds for all pure states if and only if every pair with one endpoint in $\{i,j\}$ and one in $R$ is separated by at least one further cut in $\mathcal{F}$. For example, on five parties the star $\mathcal{F}=\{\{1,\ell\}\mid 2\leq\ell\leq5\}$ has $\mu_{12}=2$ for the distinguished cut $\{1,2\}$, although the hub and the leaves have different incidence parity. Hence the parity condition of the PHEPI is sufficient but not necessary. The WTPCEI also recovers earlier inequalities with their optimal coefficients. For the edges of $C_n$ and the distinguished edge $\{1,n\}$, one finds $r_1=r_n=2$, which gives the EEI. For the edges of $K_n$, every pair is separated by $2(n-2)$ edges, which gives the case $k=2$ of Theorem~\ref{Theorem:SHMEI} and hence the TMEPI. For $n=4$, $\mathcal{F}=\{\{1,3\},\{1,4\}\}$, and the cut $\{1,2\}$, all relevant separation weights equal one, which gives Corollary~\ref{cor:four-party-triangle}.

\begin{corollary}[Cyclic Two-Party Entanglement Inequality]\label{cor:cyclic-two-party}
Let $|\psi\rangle\in\medotimes_{j=1}^{n}\mathbbm{C}^{d_j}$ be a pure $n$-partite quantum state with $n\geq3$, adopt the cyclic convention $A_{j+n}\equiv A_j$, and let $1\leq r\leq n-1$. For distinct $i,j\in[n]$, put $d\coloneqq|i-j|$ and $\kappa\coloneqq\min\{r,n-r,d,n-d\}$. Then
\begin{equation}\label{cyclic-two-party}
2\,\mathcal{E}_{A_iA_j}+\kappa\,\mathcal{I}(A_i:A_j)\leq\sum_{t=1}^{n}\mathcal{E}_{A_t\cdots A_{t+r-1}}.
\end{equation}
The coefficient two is optimal. For every realizable triple $(\mathcal{E}_{A_i},\mathcal{E}_{A_j},\mathcal{E}_{A_iA_j})$, the left-hand side is the minimum of the right-hand side over pure states with that triple.
\end{corollary}
\begin{proof}
Take $\mathcal{F}$ to be the $n$ cyclic windows of length $r$ with unit weights. A window separates two parties at cyclic separation $d'$ in exactly $2\min\{r,n-r,d',n-d'\}$ cases, so every separation weight is at least two and $W(i,j)=2\kappa$. Each $x\in\{i,j\}$ has a cyclic neighbor $\ell\notin\{i,j\}$, for which $W(x,\ell)=2$. Hence $r_i=r_j=2$. Inserting these values into Eq.~\eqref{weighted-two-party-refinement} and using $\mathcal{I}(A_i:A_R)+\mathcal{I}(A_j:A_R)=2\,\mathcal{E}_{A_iA_j}$ gives Eq.~\eqref{cyclic-two-party}. Optimality and the minimum statement follow from Theorem~\ref{theorem:weighted-two-party-cut}.
\end{proof}

\begin{remark}
For $r=2$ and adjacent parties, $\kappa=1$ and Eq.~\eqref{cyclic-two-party} becomes $\mathcal{E}_{A_k}+\mathcal{E}_{A_{k+1}}+\mathcal{E}_{A_kA_{k+1}}\leq\sum_{j=1}^{n}\mathcal{E}_{A_jA_{j+1}}$,
which strengthens the EEI by $\mathcal{I}(A_k:A_{k+1})$. Equivalently, $\Delta_{\mathrm{EEI}}\geq\mathcal{I}(A_1:A_n)$ for the edge $\{1,n\}$; this also follows from Eq.~\eqref{EEI-deficit}, since $\Delta_{\mathrm{CEI}}$ is at least its last term $\mathcal{I}(A_1\cdots A_{n-2}:A_n)\geq\mathcal{I}(A_1:A_n)$. For $r=2$, $n\geq4$, and non-adjacent parties, the EEI bound extends to every pair with correction $2\,\mathcal{I}(A_i:A_j)$. For $r\in\{1,n-1\}$, Eq.~\eqref{cyclic-two-party} reduces to the SMEI for the cut $\{i,j\}$. The weight of the mutual information grows with the separation of the pair.
\end{remark}

\begin{corollary}[Cross-Order Two-Party Entanglement Inequality]\label{cor:cross-order-two-party}
Let $|\psi\rangle\in\medotimes_{j=1}^{n}\mathbbm{C}^{d_j}$ be a pure $n$-partite quantum state with $n\geq3$, let $1\leq k\leq n-1$, and let $i,j\in[n]$ be distinct. Then
\begin{equation}\label{cross-order-two-party}
\binom{n-2}{k-1}\bigl(\mathcal{E}_{A_i}+\mathcal{E}_{A_j}+\mathcal{E}_{A_iA_j}\bigr)\leq\mathcal{E}_{\mathfrak{I}_k}.
\end{equation}
Equivalently, $2\binom{n-2}{k-1}\mathcal{E}_{A_iA_j}+\binom{n-2}{k-1}\mathcal{I}(A_i:A_j)\leq\mathcal{E}_{\mathfrak{I}_k}$. In particular, $2\binom{n-2}{k-1}\mathcal{E}_{A_iA_j}\leq\mathcal{E}_{\mathfrak{I}_k}$ at every marginal order $k$, and this coefficient is optimal. For every realizable triple $(\mathcal{E}_{A_i},\mathcal{E}_{A_j},\mathcal{E}_{A_iA_j})$, the left-hand side of Eq.~\eqref{cross-order-two-party} is the minimum of $\mathcal{E}_{\mathfrak{I}_k}$ over pure states with that triple.
\end{corollary}
\begin{proof}
Take $\mathcal{F}=\mathfrak{I}_k$ with unit weights. A $k$-party subset separates two distinct parties precisely when it contains exactly one of them, which happens for $2\binom{n-2}{k-1}$ subsets. Hence $W(i,j)=r_i=r_j=\mu_{ij}=2\binom{n-2}{k-1}$. Inserting these values into Eq.~\eqref{weighted-two-party-refinement} and using $\mathcal{I}(A_i:A_R)+\mathcal{I}(A_j:A_R)=2\,\mathcal{E}_{A_iA_j}$ and $\mathcal{I}(A_i:A_j)=\mathcal{E}_{A_i}+\mathcal{E}_{A_j}-\mathcal{E}_{A_iA_j}$ gives Eq.~\eqref{cross-order-two-party}. For optimality, place a pure entangled pair of entropy $h>0$ on $A_iA_\ell$ with $\ell\in R$ and put the remaining parties in pure states. Then $\mathcal{E}_{A_iA_j}=h$ and $\mathcal{E}_{\mathfrak{I}_k}=2\binom{n-2}{k-1}h$. The minimum statement is that of Theorem~\ref{theorem:weighted-two-party-cut}.
\end{proof}

For $k=2$, Eq.~\eqref{cross-order-two-party} strengthens Eq.~\eqref{SHMEI} for the cut $\{i,j\}$ by adding $(n-2)\,\mathcal{I}(A_i:A_j)$ to its left-hand side. For $k\neq2$, it compares a two-party cut with the total entanglement at a different marginal order, with the same optimal coefficient $2\binom{n-2}{k-1}$ as the sharp $k$-MEI at order $k$. For $k\in\{1,n-1\}$, it reduces to the SMEI for the cut $\{i,j\}$.

The same separation weights also control a distinguished one-party cut. Adjoining a party in a pure product state reduces this case to Theorem~\ref{theorem:weighted-two-party-cut}.

\begin{corollary}[Weighted One-Party Cut Entanglement Inequality]\label{cor:weighted-one-party-cut}
Let $|\psi\rangle\in\medotimes_{j=1}^{n}\mathbbm{C}^{d_j}$ be a pure $n$-partite quantum state with $n\geq2$. Fix $i\in[n]$, let $\mathcal{F}$, $w_I$, and $W$ be as in Theorem~\ref{theorem:weighted-two-party-cut}, and set $r_i\coloneqq\min_{\ell\neq i}W(i,\ell)$. Then
\begin{equation}\label{weighted-one-party-cut}
r_i\,\mathcal{E}_{A_i}\leq\sum_{I\in\mathcal{F}}w_I\,\mathcal{E}_I.
\end{equation}
The coefficient $r_i$ is optimal for the prescribed weighted family, uniformly over pure states and local dimensions. For every $h\geq0$, $r_ih$ is the minimum of $\sum_{I\in\mathcal{F}}w_I\mathcal{E}_I$ over pure states with $\mathcal{E}_{A_i}=h$, allowing arbitrary finite local dimensions.
\end{corollary}
\begin{proof}
Adjoin a party $A_{n+1}$ and consider $|\psi'\rangle\coloneqq|\psi\rangle\otimes|0\rangle_{A_{n+1}}$. For $I\in\mathcal{F}$, put $I'\coloneqq I\cup\{n+1\}$ if $i\in I$ and $I'\coloneqq I$ otherwise, and give $I'$ the weight $w_I$. Each $I'$ is a nonempty proper subset of $[n+1]$, and $\mathcal{E}_{I'}(|\psi'\rangle)=\mathcal{E}_I(|\psi\rangle)$ because $A_{n+1}$ is in a pure product state. For $\ell\in[n]\setminus\{i\}$, the cut $I'$ separates $n+1$ and $\ell$ precisely when $I$ separates $i$ and $\ell$, and it separates $i$ and $\ell$ precisely when $I$ does. Hence the separation weights $W'$ of the new family satisfy $W'(i,\ell)=W'(n+1,\ell)=W(i,\ell)$, so that $r'_i=r'_{n+1}=r_i$ and $\mu'_{i,n+1}=r_i$. Since $\mathcal{E}_{A_iA_{n+1}}(|\psi'\rangle)=\mathcal{E}_{A_i}(|\psi\rangle)$, Theorem~\ref{theorem:weighted-two-party-cut} applied to the $(n+1)$-partite state $|\psi'\rangle$ gives Eq.~\eqref{weighted-one-party-cut}. For optimality and the minimum statement, choose $\ell^\ast\neq i$ with $W(i,\ell^\ast)=r_i$, place a pure entangled pair of entropy $h>0$ on $A_iA_{\ell^\ast}$, and put the remaining parties in pure states. Then $\mathcal{E}_{A_i}=h$, and $\mathcal{E}_I=h$ precisely when $I$ separates $i$ and $\ell^\ast$, so $\sum_{I\in\mathcal{F}}w_I\mathcal{E}_I=W(i,\ell^\ast)\,h=r_i\,\mathcal{E}_{A_i}$.
\end{proof}

For the one-party cuts $\mathcal{F}=\mathfrak{I}_1$ with unit weights, $r_i=2$ and Eq.~\eqref{weighted-one-party-cut} is the EPI. For $n\geq3$ and the $n$ cyclic windows of length $r$ with unit weights, the count in the proof of Corollary~\ref{cor:cyclic-two-party} gives $W(i,\ell)=2\min\{r,n-r,d,n-d\}$ with $d\coloneqq|i-\ell|$, which is minimized by a cyclic neighbor. Hence $2\,\mathcal{E}_{A_i}\leq\sum_{t=1}^{n}\mathcal{E}_{A_t\cdots A_{t+r-1}}$ for every $1\leq r\leq n-1$; for $r=2$ this is a one-party counterpart of the EEI. For $\mathcal{F}=\mathfrak{I}_k$ with unit weights, the proof of Corollary~\ref{cor:cross-order-two-party} gives $W(i,\ell)=2\binom{n-2}{k-1}$, and hence $2\binom{n-2}{k-1}\mathcal{E}_{A_i}\leq\mathcal{E}_{\mathfrak{I}_k}$ at every marginal order. For the five-party star $\mathcal{F}=\{\{1,\ell\}\mid2\leq\ell\leq5\}$ with unit weights, the hub has $r_1=3$ and each leaf has $r_\ell=2$, so $3\,\mathcal{E}_{A_1}\leq\sum_{\ell=2}^{5}\mathcal{E}_{A_1A_\ell}$ and $2\,\mathcal{E}_{A_\ell}\leq\sum_{\ell'=2}^{5}\mathcal{E}_{A_1A_{\ell'}}$.

\begin{remark}\label{rem:weighted-one-party-refinement}
A direct proof gives a refinement of the WOPCEI in Eq.~\eqref{weighted-one-party-cut}. Fix an order on $R\coloneqq[n]\setminus\{i\}$, let $P_\ell$ and $Q_\ell$ be the sets preceding and following $\ell$, and write $D(\ell\,|\,U)\coloneqq\mathcal{E}_{U\cup\{\ell\}}-\mathcal{E}_U$ for $\ell\notin U$, as in the proof of Theorem~\ref{theorem:weighted-two-party-cut}. Then
\begin{equation}\label{weighted-one-party-refinement}
\frac12\sum_{\ell\in R}W(i,\ell)\,\mathcal{I}\big(A_i:A_\ell\mid A_{Q_\ell}\big)\leq\sum_{I\in\mathcal{F}}w_I\,\mathcal{E}_I.
\end{equation}
Indeed, set $\lambda_\ell\coloneqq\frac12\mathcal{I}(A_i:A_\ell\mid A_{Q_\ell})$. These weights are nonnegative by SSA, and the chain rule together with purity gives $\sum_{\ell\in R}\lambda_\ell=\frac12\mathcal{I}(A_i:A_R)=\mathcal{E}_{A_i}$. By complementarity, $\mathcal{E}_{\{i\}\cup Q_\ell}=\mathcal{E}_{P_\ell\cup\{\ell\}}$ and $\mathcal{E}_{\{i,\ell\}\cup Q_\ell}=\mathcal{E}_{P_\ell}$, so $2\lambda_\ell=D(\ell\,|\,P_\ell)+D(\ell\,|\,Q_\ell)$. For $V\subseteq R$ and $\ell\in V$, SSA gives $D(\ell\,|\,P_\ell)\leq D(\ell\,|\,P_\ell\cap V)$ and likewise for $Q_\ell$; both restricted sums telescope, so $\sum_{\ell\in V}\lambda_\ell\leq\mathcal{E}_V$. A cut $I$ separates $i$ from precisely the parties of the side $V\subseteq R$ of $I\,|\,\bar{I}$ not containing $i$, and $\mathcal{E}_I=\mathcal{E}_V$. Summing with weights $w_I$ and exchanging the order of summation gives Eq.~\eqref{weighted-one-party-refinement}. Since $W(i,\ell)\geq r_i$, it implies Eq.~\eqref{weighted-one-party-cut}, and it is strictly stronger whenever the separation weights $W(i,\ell)$ differ on pairs carrying nonzero $\lambda_\ell$.
\end{remark}

Merging parties extends the WOPCEI to distinguished cuts of any size, provided that no cut of the family splits the distinguished subsystem or its complement.
\begin{proposition}[Unsplit Weighted Cut Entanglement Inequality]\label{prop:unsplit-weighted-cut}
Let $|\psi\rangle\in\medotimes_{j=1}^{n}\mathbbm{C}^{d_j}$ be a pure $n$-partite quantum state with $n\geq2$, fix $1\leq k\leq n-1$ and $I_k\in\mathfrak{I}_k$, and let $\mathcal{F}$, $w_I$, and $W$ be as in Theorem~\ref{theorem:weighted-two-party-cut}. Suppose that no cut in $\mathcal{F}$ splits $I_k$, that is, $I_k\subseteq I$ or $I_k\cap I=\emptyset$ for every $I\in\mathcal{F}$, or that the same holds for $\bar{I}_k$. Set $m_{I_k}\coloneqq\min_{u\in I_k,\,v\in\bar{I}_k}W(u,v)$. Then
\begin{equation}\label{unsplit-weighted-cut}
m_{I_k}\,\mathcal{E}_{I_k}\leq\sum_{I\in\mathcal{F}}w_I\,\mathcal{E}_I,
\end{equation}
and the coefficient $m_{I_k}$ is optimal for the prescribed weighted family, uniformly over pure states and local dimensions.
\end{proposition}
\begin{proof}
Since $\mathcal{E}_{I_k}=\mathcal{E}_{\bar{I}_k}$ and $m_{I_k}$ is symmetric under exchanging $I_k$ and $\bar{I}_k$, we may assume that no cut in $\mathcal{F}$ splits $I_k$. Regard $A_{I_k}$ as a single party $\mathfrak{k}$. This gives a pure state on the $n-k+1\geq2$ parties $\{\mathfrak{k}\}\cup\bar{I}_k$. Every $I\in\mathcal{F}$ is a union of parties of this coarser system, hence a nonempty proper cut of it with the same entanglement entropy. For $v\in\bar{I}_k$, a cut $I\in\mathcal{F}$ separates $\mathfrak{k}$ and $v$ precisely when it separates $u$ and $v$ for one, equivalently every, $u\in I_k$. The separation weights $W'$ of the coarser system therefore satisfy $W'(\mathfrak{k},v)=W(u,v)$ for all $u\in I_k$, and $\min_{v\in\bar{I}_k}W'(\mathfrak{k},v)=m_{I_k}$. Corollary~\ref{cor:weighted-one-party-cut}, applied to the party $\mathfrak{k}$, gives Eq.~\eqref{unsplit-weighted-cut}. For optimality, choose $u\in I_k$ and $v\in\bar{I}_k$ with $W(u,v)=m_{I_k}$, place a pure entangled pair of entropy $h>0$ on $A_uA_v$, and put the remaining parties in pure states. Then $\mathcal{E}_{I_k}=h$ and $\sum_{I\in\mathcal{F}}w_I\mathcal{E}_I=W(u,v)\,h=m_{I_k}\mathcal{E}_{I_k}$.
\end{proof}

The WTPCEI controls one distinguished cut. We next bound the combined entanglement of several cuts that share a party. For ordinary stars, a Hamiltonian decomposition of the complete graph gives the following bound.

\begin{theorem}[Star Entanglement Inequality]
\label{Theorem:SEI}
Let $|\psi\rangle\in\medotimes_{j=1}^{n}\mathbbm{C}^{d_j}$ be a pure $n$-partite quantum state with $n\geq3$. Fix $k\in[n]$ and let $N_k\subset[n]\setminus\{k\}$ with $|N_k|=\lfloor n/2\rfloor$. Then
\begin{equation}\label{SEI}
2\sum_{l\in N_k}\mathcal{E}_{A_kA_l}\leq\sum_{1\leq i<j\leq n}
\mathcal{E}_{A_iA_j}.
\end{equation}
\end{theorem}
\begin{proof}
Suppose first that $n=2r+1$ is odd. By the Walecki construction \cite{Alspach08}, $K_{2r+1}$ admits an edge decomposition into $r$ Hamiltonian cycles, i.e., $K_{2r+1}=\mathfrak{C}^{(1)}\sqcup\cdots\sqcup\mathfrak{C}^{(r)}$. Fix the vertex $k$. Choose one edge incident with $k$ from each Hamiltonian cycle $\mathfrak{C}^{(s)}$. Applying the EEI to each $\mathfrak{C}^{(s)}$, with the chosen edge as the distinguished edge, and summing over $s=1,\ldots,r$ gives
\begin{equation*}
2\sum_{s=1}^{r}\mathcal{E}_{e_s}\leq
\sum_{1\leq i<j\leq n}\mathcal{E}_{A_iA_j}.
\end{equation*}
The selected edges are $r$ distinct edges incident with $k$. Starting from such a decomposition and applying a permutation of the vertices other than $k$, they may be chosen to coincide with $\{\{k,l\}\mid l\in N_k\}$ for any prescribed subset $N_k$ of cardinality $r$. Now let $n=2r$ be even. In this case, the Walecki construction gives an edge decomposition $K_{2r}=M\sqcup\mathfrak{C}^{(1)}\sqcup\cdots\sqcup\mathfrak{C}^{(r-1)}$, where $M$ is a perfect matching and $\mathfrak{C}^{(1)},\ldots,\mathfrak{C}^{(r-1)}$ are Hamiltonian cycles. Let $e_0$ be the unique edge of $M$ incident with $k$, and choose one edge $e_s$ incident with $k$ from each $\mathfrak{C}^{(s)}$. Since a perfect matching is $1$-regular, the PGEPI in Corollary~\ref{cor:parity-graph} gives
\begin{equation*}
2\,\mathcal{E}_{e_0}\leq\sum_{e\in M}\mathcal{E}_{e}.
\end{equation*}
Applying the EEI to every Hamiltonian cycle gives
\begin{equation*}
2\,\mathcal{E}_{e_s}\leq\sum_{e\in E(\mathfrak{C}^{(s)})}\mathcal{E}_{e},
\qquad s=1,\ldots,r-1.
\end{equation*}
Summing these inequalities and using the edge decomposition of $K_{2r}$ yields
\begin{equation*}
2\sum_{s=0}^{r-1}\mathcal{E}_{e_s} \leq
\sum_{1\leq i<j\leq n}\mathcal{E}_{A_iA_j}.
\end{equation*}
The selected edges are $r$ distinct edges incident with $k$, and again a permutation fixing $k$ allows them to be identified with $\{\{k,l\}\mid l\in N_k\}$ for any prescribed subset $N_k$ of cardinality $r$.
\end{proof}

\begin{remark}
For odd $n$, Theorem~\ref{Theorem:SEI} reduces to the star inequality obtained directly from the Hamiltonian decomposition of $K_n$. The even case additionally uses the perfect-matching component of the Walecki decomposition. Thus, the star inequality holds uniformly for both parities of $n$.
\end{remark}

The collective construction extends to uniform hyperstars. Here a $k$-uniform hyperstar means a collection of $k$-party subsystem sets sharing a specified vertex; the remaining vertices may overlap. Averaged SSA on the complementary subsystem gives the following extension of the star inequality, using the sharp marginal bound in Theorem~\ref{Theorem:SHMEI} and SSA as inputs.

\begin{theorem}[Hyperstar Entanglement Inequality]
\label{Theorem:HSEI}
Let $|\psi\rangle\in\medotimes_{j=1}^{n}\mathbbm{C}^{d_j}$ be a pure $n$-partite state with $n\geq4$, and let $2\leq k\leq\lfloor n/2\rfloor$. Fix a vertex $v\in[n]$ and a partial hyperstar $\mathscr{S}\subseteq\mathscr{S}_v^{(k)}$ with $|\mathscr{S}|=q$, where $1\leq q\leq\binom{n-1}{k-1}$. Then
\begin{equation}\label{HSEI}
c_{n,k,q}\sum_{I_k\in\mathscr{S}}\mathcal{E}_{I_k}\leq
\mathcal{E}_{\mathfrak{I}_k}, \qquad
c_{n,k,q}=\max\Bigl\{2,\frac{2}{q}\binom{n-2}{k-1}\Bigr\}.
\end{equation}
The coefficient is optimal as a function of $n,k,q$, uniformly over pure states, local dimensions, and the choice of the partial hyperstar $\mathscr{S}$.
\end{theorem}
\begin{proof}
Summing the $q$ instances of Theorem~\ref{Theorem:SHMEI} indexed by $\mathscr{S}$ gives
\begin{equation*}
2\binom{n-2}{k-1}\sum_{I_k\in\mathscr{S}}\mathcal{E}_{I_k} \leq q\,\mathcal{E}_{\mathfrak{I}_k}.
\end{equation*}
It remains to prove the bound with coefficient two. Consider the mixed marginal on the $n-1$ parties other than $A_v$, and let
\begin{equation*}
M_r\coloneqq\binom{n-1}{r}^{\!\!-1}\!\!\sum_{\substack{I_r\subset[n]\setminus\{v\}\\|I_r|=r}}
\mathcal{S}(\varrho_{I_r}),\qquad 0\leq r\leq n-1,
\end{equation*}
be its average $r$-party entropy, so that $M_0=0$. The SSA summation in the proof of Lemma~\ref{lem:average-marginal-hierarchy} applies verbatim with $n$ replaced by $n-1$ and does not use purity, so it gives concavity of $M_r$ without requiring purity of this marginal, i.e.,  $2M_r\geq M_{r-1}+M_{r+1}$ for $1\leq r\leq n-2$. Since $k\leq n-k$, we obtain
\begin{equation*}
M_k\geq\frac{k}{n-k}M_{n-k}.
\end{equation*}
Binomial counting and global purity therefore give
\begin{equation*}
\sum_{\substack{I_k\in\mathfrak{I}_k\\v\notin I_k}}
\mathcal{E}_{I_k}=\binom{n-1}{k}M_k\geq\binom{n-1}{k}\frac{k}{n-k}M_{n-k}
=\binom{n-1}{k-1}M_{n-k}=\sum_{\substack{I_k\in\mathfrak{I}_k\\v\in I_k}}
\mathcal{E}_{I_k}.
\end{equation*}
Thus the full hyperstar carries at most one half of $\mathcal{E}_{\mathfrak{I}_k}$. Non-negativity gives the same bound for its subfamily $\mathscr{S}$. Combining the two coefficients proves Eq.~\eqref{HSEI}. To prove optimality, set $L\coloneqq\binom{n-2}{k-1}$ and fix $u\neq v$. Place a pure entangled pair of entropy $h>0$ on $A_vA_u$, with all remaining parties in pure product states. Define $\mathscr{T}_u\coloneqq\{I_k\in\mathscr{S}_v^{(k)}\mid u\notin I_k\}$. Then $|\mathscr{T}_u|=L$, $\mathcal{E}_{\mathfrak{I}_k}=2Lh$, and $\sum_{I_k\in\mathscr{S}}\mathcal{E}_{I_k}= h\,|\mathscr{S}\cap\mathscr{T}_u|$. If $q\leq L$, choose $\mathscr{S}\subseteq\mathscr{T}_u$ with $|\mathscr{S}|=q$. If $q\geq L$, choose $\mathscr{T}_u\subseteq\mathscr{S} \subseteq\mathscr{S}_v^{(k)}$ with $|\mathscr{S}|=q$. In either case,
\begin{equation*}
\frac{\mathcal{E}_{\mathfrak{I}_k}}
{\sum_{I_k\in\mathscr{S}}\mathcal{E}_{I_k}}=
\frac{2L}{\min\{q,L\}}=\max\left\{2,\frac{2L}{q}\right\}
=c_{n,k,q}.
\end{equation*}
Thus, for every admissible $n,k,q$, a pure state and a partial hyperstar of size $q$ attain equality with a nonzero selected sum. No larger coefficient depending only on $n,k,q$ can hold uniformly.
\end{proof}

\begin{remark}[Sharp collective star refinement]\label{rem:sharp-star-refinement}
At marginal order two, Theorem~\ref{Theorem:HSEI} gives
\begin{equation}
c_{n,q}\sum_{l\in N_v}\mathcal{E}_{A_vA_l}\leq \mathcal{E}_{\mathfrak{I}_2}, \qquad
c_{n,q}=\max\Bigl\{2,\frac{2(n-2)}{q}\Bigr\},
\end{equation}
where $n\geq4$, $v\in[n]$, and $N_v\subseteq[n]\setminus\{v\}$ has cardinality $q\geq1$. The coefficient is sharp for every prescribed $N_v$. To attain equality, place a pure entangled pair of entropy $h>0$ on $A_vA_u$ and tensor it with pure states on the remaining parties. Choose $u\notin N_v\cup\{v\}$ if $q\leq n-2$, and choose any $u\in N_v$ if $q=n-1$. The selected star sum is then $\min\{q,n-2\}h$, while $\mathcal{E}_{\mathfrak{I}_2}=2(n-2)h$. For $q=\lfloor n/2\rfloor$ and $n\geq5$, one has $c_{n,q}>2$. Hence this bound strictly improves the SEI, and equality in Eq.~\eqref{SEI} forces $\mathcal{E}_{\mathfrak{I}_2}=0$; by Lemma~\ref{lem:average-marginal-hierarchy}, the state is then fully separable.
\end{remark}

\begin{remark}\label{rem:hyperstar-range-sharpness}
For $n=2k$, the full-hyperstar comparison is an identity, since complementation pairs every $k$-subset containing $v$ with one not containing $v$. For $k\geq3$, optimality of the universal coefficient does not assert optimality for each prescribed family: the proof selects families attaining the bound, while a particular incidence pattern may allow a stronger inequality. The range $k\leq\lfloor n/2\rfloor$ is essential for the coefficient-two full-hyperstar statement. For example, on four parties put an entangled pair of entropy $h$ on $A_2A_3$ and make $A_1,A_4$ pure. The three three-party hyperedges containing vertex $1$ have entropies $0,h,h$, whereas $\mathcal{E}_{\mathfrak{I}_3}=2h$. A coefficient of two for that full hyperstar would therefore be false.
\end{remark}

Finally, the complete $k$-uniform hypergraph provides a direct graph-theoretic interpretation of the GEPI. Its hyperedges are the elements of $\mathfrak{I}_k$, and each vertex belongs to exactly $\binom{n-1}{k-1}$ hyperedges. Hence all vertex degrees have the same parity, and the PHEPI in Theorem~\ref{Theorem:parity-hypergraph} gives $2\,\mathcal{E}_{I_k}\leq\mathcal{E}_{\mathfrak{I}_k}$. Thus, the GEPI also arises directly as the polygon inequality of the complete $k$-uniform hypergraph, while the sharp $k$-MEI in Theorem~\ref{Theorem:SHMEI} provides the stronger quantitative bound (see Eq.~\eqref{SHMEI}).

\section{Examples and Applications}\label{sec:applications}

We illustrate sharpness, equality, and the role of subsystem geometry before turning to tensor networks and recoverability.

\subsection{Representative States and Sharpness}
\label{subsec:examples}

We write $H_2(p)=-p\log_2p-(1-p)\log_2(1-p)$. Numerical subscripts abbreviate the corresponding sets of parties.

The following example examines how the CEI deficit in Theorem~\ref{Theorem:CEI} depends on the ordering of the parties.
\begin{example}\label{ex:four-qubit-chain}
Table~\ref{tab:chain-examples} compares four-qubit states with identical one-party entropies. Here $|\Phi^+\rangle=(|00\rangle+|11\rangle)/\sqrt2$, $|\mathrm{GHZ}_4\rangle=(|0000\rangle+|1111\rangle)/\sqrt2$, and $|\mathrm{Cl}_4\rangle=\mathrm{CZ}_{12}\mathrm{CZ}_{23} \mathrm{CZ}_{34}|+\rangle^{\otimes4}$~\cite{BR01},
where $\mathrm{CZ}_{ij}$ is the controlled-$Z$ gate acting on qubits $i$ and $j$. The adjacent Bell-pair product saturates the CEI, whereas the crossed product has deficit four. Ordering the latter as $A_1,A_3,A_2,A_4$ makes its deficit zero. Thus the deficit depends on the ordering and can be positive for a state that is separable across a bipartition.
\begin{table}[h]
\centering
\begin{tabular}{lcccc}
\hline\hline
State & $\mathcal{E}_{12}$ & $\mathcal{E}_{23}$
& $\mathcal{E}_{34}$ & $\Delta_{\mathrm{CEI}}$\\
\hline
$|\mathrm{GHZ}_4\rangle$ & $1$ & $1$ & $1$ & $1$\\
$|\Phi^+\rangle_{12}|\Phi^+\rangle_{34}$
& $0$ & $2$ & $0$ & $0$\\
$|\Phi^+\rangle_{13}|\Phi^+\rangle_{24}$
& $2$ & $2$ & $2$ & $4$\\
$|\mathrm{Cl}_4\rangle$ & $1$ & $2$ & $1$ & $2$\\
\hline\hline
\end{tabular}
\caption{Four-qubit states with $\mathcal{E}_{A_j}=1$ for every party. The chain deficit is $\Delta_{\mathrm{CEI}}= \mathcal{E}_{12}+\mathcal{E}_{23}+\mathcal{E}_{34}-2$. All values are in bits.}
\label{tab:chain-examples}
\end{table}
\end{example}

The following example compares saturation of the average hierarchy in Lemma~\ref{lem:average-marginal-hierarchy} and the average upper bound in Corollary~\ref{cor:average-k-marginal-entanglement}. It also distinguishes the prefix equality conditions in Theorem~\ref{Theorem:CEI} from local three-party tests.
\begin{example}
\label{ex:marginal.profiles-ghz.local.cmi-AME.average.bound}
For $n\geq3$ and $0<p<1$, the state $|\mathrm{GHZ}_n(p)\rangle=\sqrt p\,|0\rangle^{\otimes n}+\sqrt{1-p}\,|1\rangle^{\otimes n}$ has entropy $h=H_2(p)$ on every nonempty proper subsystem. Hence $\overline{\mathcal{E}}_r=h$ at every proper order, while $\Delta_{\mathrm{CEI}}=h>0$. For $n\geq4$, all consecutive triples satisfy $\mathcal{I}(A_j:A_{j+2}\mid A_{j+1})=0$. Nevertheless, the last prefix conditional mutual information term equals $h$ and all earlier terms vanish. Local three-party tests therefore do not replace the prefix conditions in Eq.~\eqref{Markov-chain}. In contrast, an $\mathrm{AME}(n,d)$ state has $\overline{\mathcal{E}}_r=\min\{r,n-r\}\log_2d$ and saturates the average upper bound at every order. For example~\cite{GALRZ15},
\begin{equation*}
|\mathrm{AME}_{4,3}\rangle
=\frac13\sum_{i,j\in\mathbbm{Z}_3}|i,j,i+j,i+2j\rangle,
\end{equation*}
with additions modulo three. Its one- and two-party marginals are maximally mixed, giving $\overline{\mathcal{E}}_1=\log_2 3$ and $\overline{\mathcal{E}}_2=2\log_2 3$.
\end{example}

Independent pure bonds illustrate equality and strictness in Corollary~\ref{cor:k-MEI}, Theorem~\ref{Theorem:SHMEI}, and the powered bounds in Proposition~\ref{prop:GEPI-alpha}. The same construction compares saturation of the hyperchain and cyclic inequalities in Theorems~\ref{Theorem:HCEI} and~\ref{Theorem:EEI} and Corollary~\ref{Theorem:HEEI}.
\begin{example}
\label{ex:finite.range.bonds-sharp.and.weak.marginals-cyclic.bond.saturation}
Let $\mathscr{B}$ be a family of pairs carrying independent pure bonds of entropies $h_{ab}>0$. Bonds meeting at a party occupy separate local tensor factors, possibly embedded by a local isometry; unused factors are pure. Additivity gives
\begin{equation*}
\mathcal{E}_I=\sum_{\substack{\{a,b\}\in\mathscr{B}\\
|\{a,b\}\cap I|=1}}h_{ab}, \qquad\qquad\qquad
\Delta_m=2\sum_{\substack{\{a,b\}\in\mathscr{B}\\
a<b,\ b-a>m}}h_{ab}.
\end{equation*}
Each bond of span greater than $m$ contributes $2h_{ab}$ to exactly one prefix conditional mutual information term. Thus HCEI equality within this family is equivalent to all bond spans being at most $m$. Counting the marginals crossing each bond shows that the sharp $k$-marginal inequality is saturated precisely when every bond crosses the distinguished cut. For a single bond of entropy $h$ crossing $I_k\mid\bar{I}_k$,
\begin{equation}\label{single-bond-sharpness}
\mathcal{E}_{I_k}=h, \qquad\qquad
\mathcal{E}_{\mathfrak{I}_k}
=2\binom{n-2}{k-1}h, \qquad\qquad
\overline{\mathcal{E}}_k
=\frac{2k(n-k)}{n(n-1)}h.
\end{equation}
This state also saturates the MEI, SMEI, and the EPI at either endpoint, as well as their powered forms and the sharp power bound. Its average profile is strictly increasing up to the middle order. For a single bond with endpoints $a<b$, put $d=b-a$ and let $I_r^{(j)}=\{j,\ldots,j+r-1\}$ be cyclic windows. Counting those separating the endpoints gives
\begin{equation}\label{cyclic-bond-count}
\sum_{j=1}^{n}\mathcal{E}_{I_r^{(j)}}
=2\min\{r,n-r,d,n-d\}\,h.
\end{equation}
A cyclically adjacent bond therefore saturates the EEI and HEEI whenever the distinguished window separates its endpoints. In particular, a bond on $A_1A_n$ saturates these cyclic bounds but has $\Delta_m=2h$ throughout the hyperchain range. For comparison, equality in the weaker $k$-MEI at $2\leq k\leq\lfloor n/2\rfloor$ forces $\mathcal{E}_{\mathfrak{I}_k}=0$ by the sharp bound. The hierarchy then forces every one-party entropy to vanish, so the state is a product of one-party states. The same argument applies to the GEPI at $2\leq k\leq n-2$, including the TMEPI for $n\geq4$.
\end{example}

The following example gives nonzero equality cases for Lemma~\ref{lem:symmetric-difference} and the PHEPI in Theorem~\ref{Theorem:parity-hypergraph}, covering both even and odd incidence.
\begin{example}\label{ex:parity-equalities}
Put a pure entangled pair of entropy $h>0$ on $A_1A_2$ and make the other parties pure. The family $\{12,234,134\}$ has even incidence at every party and entropy profile $(0,h,h)$. Since $12\triangle234=134$, it saturates both the symmetric-difference triangle and the polygon bounds for the nonzero terms. For odd incidence, take $n=9$ and the family $\{123,456,789\}$, and put the pair on $A_3A_4$. The profile is $(h,h,0)$, again giving nonzero polygon equalities.
\end{example}

The following example illustrates the separation weights in the WTPCEI (Theorem~\ref{theorem:weighted-two-party-cut}), the gap between its refined and polygon forms, and saturation of the refined bound by genuinely multipartite entangled states.
\begin{example}\label{ex:weighted-two-party-ghz}
Take $n=5$ and $\mathcal{F}=\{\{1,\ell\}\mid 2\leq\ell\leq5\}$ with $w_{12}=2$ and $w_{13}=w_{14}=w_{15}=1$. For the distinguished cut $\{1,2\}$, the separation weights are $W(1,2)=3$, $W(1,\ell)=4$, and $W(2,\ell)=3$ for $\ell\in\{3,4,5\}$. Thus $r_1=4$, $r_2=3$, and $\mu_{12}=3$. Put $|\mathrm{GHZ}_3\rangle=(|000\rangle+|111\rangle)/\sqrt{2}$ on $A_1A_2A_3$ and let $A_4,A_5$ be in $|0\rangle$. Every cut in $\mathcal{F}$ has entropy one bit, so $\sum_{I\in\mathcal{F}}w_I\mathcal{E}_I=5$. With $A_R=A_3A_4A_5$, the mutual information terms $\mathcal{I}(A_1:A_2)$, $\mathcal{I}(A_1:A_R)$, and $\mathcal{I}(A_2:A_R)$ also equal one bit. The left-hand side of Eq.~\eqref{weighted-two-party-refinement} is therefore $3+4+3=10$, and the refined bound is saturated. The polygon form is strict, $\mu_{12}\mathcal{E}_{12}=3<5$, because passing from Eq.~\eqref{weighted-two-party-refinement} to Eq.~\eqref{weighted-two-party-cut} discards $W(1,2)\,\mathcal{I}(A_1:A_2)$ and replaces $r_1$ by $\mu_{12}$. The same entropy values, and hence the same equality, occur for the genuinely five-party entangled state $|\mathrm{GHZ}_5\rangle$; the minimizers are thus not restricted to products of bipartite bonds. Finally, a single Bell pair on $A_2A_3$ with pure spectators has $\mathcal{E}_{12}=1$ and weighted sum $W(2,3)=3=\mu_{12}$, so the polygon coefficient is attained.
\end{example}

The following example illustrates saturation of both coefficient regimes in the HSEI (Theorem~\ref{Theorem:HSEI}).
\begin{example}\label{ex:hyperstar-saturation}
On six parties, put a pure entangled pair of entropy $h$ on $A_1A_6$ and make the remaining parties pure. For $\mathscr{S}=\{123,124,135,145\}$, the selected sum is $4h$ and $\mathcal{E}_{\mathfrak{I}_3}=12h$, so the coefficient $c_{6,3,4}=3$ is attained. For the coefficient-two regime, use seven parties with the pair on $A_1A_7$ and pure spectator states. Of the fifteen triples containing vertex $1$, ten have entropy $h$ and five have entropy zero. The full hyperstar sum is therefore $10h$, whereas $\mathcal{E}_{\mathfrak{I}_3}=20h$.
\end{example}

\subsection{Tensor Networks and Saturation}
\label{subsec:tensor-networks}

An open-boundary Matrix Product State (MPS) has the form
\begin{equation}
|\psi\rangle=\sum_{s_1,\ldots,s_n}
M_1^{s_1}M_2^{s_2}\cdots M_n^{s_n}|s_1\cdots s_n\rangle,
\end{equation}
where $M_j^{s_j}$ is a $D_{j-1}\times D_j$ matrix and $D_0=D_n=1$. The dimensions $D_j$ bound the Schmidt ranks across prefix cuts. A contiguous interval $A_a\cdots A_b$ has at most two exposed virtual bonds, giving the standard rank and entropy bounds \cite{PVWC07,CPSV21}
\begin{equation}
\operatorname{rank}\varrho_{A_a\cdots A_b}\leq D_{a-1}D_b,
\qquad
\mathcal{E}_{A_a\cdots A_b}\leq\log_2(D_{a-1}D_b).
\end{equation}
These statements are independent of the CEI. Combining the two-party case with the exact CEI deficit gives the following capacity condition:
\begin{equation}
\sum_{j=2}^{n-1}\mathcal{E}_{A_j}+\Delta_{\mathrm{CEI}}
\leq\log_2D_1+2\sum_{j=2}^{n-2}\log_2D_j+\log_2D_{n-1}.
\end{equation}
In particular, if $D_j\leq D$, then
\begin{equation}
D\geq2^{\left(\sum_{j=2}^{n-1}\mathcal{E}_{A_j}
+\Delta_{\mathrm{CEI}}\right)/(2(n-2))}.
\end{equation}
This is a necessary condition for an exact MPS representation. Dropping the deficit recovers a bound that also follows by summing the standard one-site rank bounds. The deficit retains the additional information in the two-party entropies; neither bound alone certifies an accurate approximation of the state.

The same reasoning applies to the hyperchain. If $\Delta_m$ denotes its pure-state deficit, then
\begin{equation}
\sum_{j=1}^{n-m-1}\mathcal{E}_{A_{j+1}\cdots A_{j+m}}+\Delta_m
\leq\sum_{j=1}^{n-m}\log_2(D_{j-1}D_{j+m}).
\end{equation}
Increasing the overlap size therefore gives constraints at different block scales. Finite MPS bond dimension by itself does not imply equality in these constraints or a vanishing Markov deficit: the GHZ and cluster examples in Table~\ref{tab:chain-examples} already have small bond dimension and positive CEI deficit.

The direct entropy proofs in Sec.~\ref{sec:Ent.Ineq.} also connect the chain inequalities to existing information-theoretic constructions. In entropy notation, the CEI and HCEI correspond, respectively, to one-site and consecutive-block choices of conditioning subsystems in the Markov entropy decomposition of Ref.~\cite{PH11}. This connection gives a structural interpretation of their exact deficits: vanishing conditional mutual information is the equality condition for SSA characterized in Ref.~\cite{HJPW04}. The following pure-state chain characterization is a specialization of these equality results and the tree structure theorem in Ref.~\cite[Appendix A, Theorem 3]{PH11}. We give a direct inductive proof in the present entanglement notation.

\begin{proposition}[Pure-state saturation of the CEI \cite{PH11,HJPW04}]\label{prop:pure-chain-saturation}
A pure state saturates the CEI if and only if there are auxiliary Hilbert spaces $\mathcal{H}_{j,L},\mathcal{H}_{j,R}$, local isometries $V_j:\mathcal{H}_{j,L}\otimes\mathcal{H}_{j,R}\to\mathcal{H}_j$, and normalized bipartite pure states $|\omega_j\rangle$ such that
\begin{equation}\label{pure-chain-bonds}
|\psi\rangle=\Bigl(\bigotimes_{j=1}^{n}V_j\Bigr)
\bigotimes_{j=1}^{n-1}|\omega_j\rangle_{j,R\,;\,j+1,L}.
\end{equation}
The end spaces $\mathcal{H}_{1,L}$ and $\mathcal{H}_{n,R}$ may be taken one-dimensional. The auxiliary spaces may be restricted to the Schmidt supports of the bond states, so that the range of $V_j$ is $\operatorname{supp}\varrho_{A_j}$.
\end{proposition}
\begin{proof}
Suppose first that $\Delta_{\mathrm{CEI}}=0$. The last conditional mutual information in Eq.~\eqref{CEI-deficit} vanishes. For the pure tripartition $P=A_1\cdots A_{n-2}$, $B=A_{n-1}$, and $C=A_n$, purity gives $\mathcal{I}(P:C\mid B)=\mathcal{I}(P:C)$, hence $\varrho_{PC}=\varrho_P\otimes\varrho_C$. Moreover, purity of the state on $PBC$ implies
\begin{equation*}
\operatorname{rank}\varrho_B=\operatorname{rank}\varrho_{PC}
=\operatorname{rank}\varrho_P\,\operatorname{rank}\varrho_C.
\end{equation*}
A minimal purification of a product state factorizes. Accordingly, on the support of $B$ there is an identification $B=B_L\otimes B_R$ under which
\begin{equation*}
|\psi\rangle=|\chi\rangle_{P B_L}\otimes|\omega_{n-1}\rangle_{B_R C},
\end{equation*}
up to an isometry into $\mathcal{H}_{n-1}$. Let $h=\mathcal{S}(\varrho_C)$. The factorization above gives $\mathcal{S}(\varrho_{A_{n-2}B})=\mathcal{S}(\varrho_{A_{n-2}B_L})+h$, $\mathcal{S}(\varrho_B)=\mathcal{S}(\varrho_{B_L})+h$, and $\mathcal{S}(\varrho_{BC})=\mathcal{S}(\varrho_{B_L})$. Let $\Delta_{\mathrm{CEI}}(\chi)$ denote the chain deficit of the pure state $|\chi\rangle$ on the shorter chain $A_1,\ldots,A_{n-2},B_L$. The entropy identities above imply
\begin{equation*}
\Delta_{\mathrm{CEI}}-\Delta_{\mathrm{CEI}}(\chi)
=h+\mathcal{S}(\varrho_{B_L})-\bigl(\mathcal{S}(\varrho_{B_L})+h\bigr)=0.
\end{equation*}
Hence $\Delta_{\mathrm{CEI}}(\chi)=0$. Repeating this argument reduces the chain to a pure two-party state, which itself forms a single bond. Induction therefore gives Eq.~\eqref{pure-chain-bonds}. Conversely, let $h_j$ be the entanglement entropy of $|\omega_j\rangle$, with $h_0=h_n=0$. In the state of Eq.~\eqref{pure-chain-bonds},
\begin{equation}\label{bond-state-entropies}
\mathcal{E}_{A_j}=h_{j-1}+h_j, \qquad
\mathcal{E}_{A_jA_{j+1}}=h_{j-1}+h_{j+1}.
\end{equation}
Local isometries preserve the nonzero marginal spectra. Substituting these expressions into the deficit makes it zero.
\end{proof}

Thus exact saturation has a concrete virtual-bond interpretation: all correlation can be represented by independent pure bonds between adjacent sites, embedded isometrically into the physical sites. This is a special class of MPS, rather than a property of arbitrary MPS. Nonunitary physical maps used in a general tensor network need not preserve this structure.

\begin{corollary}[Qubit chains]\label{cor:qubit-chain-saturation}
A pure qubit state saturates the CEI if and only if it is a product of pure one-qubit states and pure two-qubit states supported on disjoint adjacent pairs.
\end{corollary}
\begin{proof}
Let $r_j$ be the Schmidt rank of $|\omega_j\rangle$ and set $r_0=r_n=1$. The support form in Proposition~\ref{prop:pure-chain-saturation} gives $\operatorname{rank}\varrho_{A_j}=r_{j-1}r_j\leq2$. Two nontrivial bonds therefore cannot meet at a qubit. The nontrivial bonds form a matching of the path, and the remaining local isometries can be absorbed into the pure one- and two-qubit factors. The converse follows from Eq.~\eqref{bond-state-entropies}.
\end{proof}

For higher local dimension, adjacent nontrivial bonds are possible. For example, internal sites of dimension four can each carry the two qubit halves of two neighboring Bell bonds. In general, every exact saturator obeys $r_{j-1}r_j\leq d_j$. Saturation therefore constrains how the physical dimension must accommodate the independent virtual factors.

For overlaps of at least two sites, HCEI equality no longer requires independent adjacent bonds. The following example uses open-chain cluster states to compare equality and strictness in the CEI and HCEI (Theorems~\ref{Theorem:CEI} and~\ref{Theorem:HCEI}) and the HEEI (Corollary~\ref{Theorem:HEEI}).
\begin{example}\label{ex:cluster-hyperchain}
For the open cluster state~\cite{BR01}
\begin{equation*}
|\mathrm{Cl}_n\rangle
=\prod_{j=1}^{n-1}\mathrm{CZ}_{j,j+1}
|+\rangle^{\otimes n},\qquad n\geq4,
\end{equation*}
a subsystem entropy equals the binary rank of its graph cut matrix~\cite{HEB04}. Every one-party entropy is one, while a proper contiguous interval of at least two sites has entropy one if it touches an endpoint and two otherwise. Consequently, $\Delta_{\mathrm{CEI}}=n-2$ and $\Delta_m=0$ for $2\leq m\leq n-2$. The connected path graph makes the state genuinely multipartite entangled, so HCEI equality allows states beyond independent adjacent bonds. For $|\mathrm{Cl}_6\rangle$, the cyclic three-party entropies are $(1,2,2,1,2,2)$; hence every corresponding HEEI is strict despite HCEI equality.
\end{example}

For $n\geq4$, the cycle inequality admits a further decomposition that directly controls its equality and near-equality regimes. For the distinguished edge $\{1,n\}$, write
\begin{equation}
\Delta_{\mathrm{EEI}}\coloneqq
\sum_{j=1}^{n}\mathcal{E}_{A_jA_{j+1}}-2\,\mathcal{E}_{A_1A_n}, \qquad A_{n+1}\equiv A_1.
\end{equation}
Define the block-entropy sum
\begin{equation}\label{EEI-block-entropy}
\Omega\coloneqq\mathcal{E}_{A_1A_2}+\sum_{j=3}^{n-2}\mathcal{E}_{A_j}+\mathcal{E}_{A_{n-1}A_n}.
\end{equation}
The middle sum is empty when $n=4$. Define the interior deficit by
\begin{equation}
\widehat\Delta\coloneqq\sum_{j=2}^{n-2}\mathcal{S}(\varrho_{A_jA_{j+1}})
-\sum_{j=3}^{n-2}\mathcal{S}(\varrho_{A_j})
-\mathcal{S}(\varrho_{A_2\cdots A_{n-1}}).
\end{equation}
For $n\geq5$, applying the mixed-state chain bound in Eq.~\eqref{mixed-chain-entropy} to the interior marginal on $A_2\cdots A_{n-1}$ gives $\widehat\Delta\geq0$. For $n=4$, the expression reduces identically to
\begin{equation*}
\widehat\Delta=\mathcal{S}(\varrho_{A_2A_3})-\mathcal{S}(\varrho_{A_2A_3})=0.
\end{equation*}
Thus $\widehat\Delta\geq0$ for every $n\geq4$. Purity, together with the definitions, yields the exact identity
\begin{equation}\label{EEI-stability-decomposition}
\Delta_{\mathrm{EEI}}=\Omega+\widehat\Delta\geq\Omega.
\end{equation}
The following consequence strengthens the equality characterization to a quantitative approximation statement.

\begin{corollary}[Exact and approximate saturation of EEI]\label{cor:cycle-edge-saturation}
Let $n\geq4$. The EEI for the distinguished edge $\{1,n\}$ is saturated if and only if the pure state factorizes as
\begin{equation}\label{EEI-saturation-state}
|\psi\rangle=|\eta\rangle_{A_1A_2}\otimes \Bigl(\bigotimes_{j=3}^{n-2}|\alpha_j\rangle_{A_j}\Bigr) \otimes|\zeta\rangle_{A_{n-1}A_n},
\end{equation}
where the middle product is absent when $n=4$. The two end-pair states may be entangled. Moreover, for every pure state $|\psi\rangle$ there exists a normalized state $|\Phi\rangle$ of the displayed product form such that
\begin{equation}\label{EEI-stability-overlap}
|\langle\Phi|\psi\rangle|^2
\geq2^{-\Omega}\geq2^{-\Delta_{\mathrm{EEI}}},
\end{equation}
and therefore
\begin{equation}\label{EEI-stability-distance}
\frac{1}{2}\big\||\psi\rangle\langle\psi|-|\Phi\rangle\langle\Phi|\big\|_1
\leq\sqrt{1-2^{-\Delta_{\mathrm{EEI}}}}
\leq\sqrt{\ln2\;\Delta_{\mathrm{EEI}}}.
\end{equation}
The bounds hold in arbitrary finite local dimensions. Other distinguished edges are obtained by cyclic relabeling.
\end{corollary}
\begin{proof}
If $\Delta_{\mathrm{EEI}}=0$, Eq.~\eqref{EEI-stability-decomposition} forces $\Omega=0$. Every block marginal in Eq.~\eqref{EEI-block-entropy} is then pure, so the global pure state factorizes over these blocks. This is exactly Eq.~\eqref{EEI-saturation-state}. Conversely, a state of that form has $\Omega=\widehat\Delta=0$, and hence saturates the EEI. For the quantitative statement, let $\mathfrak P$ denote the partition into $\{1,2\}$, the singleton blocks $\{3\},\ldots,\{n-2\}$, and $\{n-1,n\}$. Put $\sigma=\medotimes_{B\in\mathfrak P}\varrho_B$. The vector $|\psi\rangle$ lies in the tensor product of the marginal supports. Consequently, with the logarithm restricted to that support,
\begin{equation*}
-\langle\psi|\log_2\sigma|\psi\rangle
=\sum_{B\in\mathfrak P}\mathcal{S}(\varrho_B)=\Omega.
\end{equation*}
Concavity of the scalar logarithm, applied to the spectral decomposition of $\sigma$, gives $\langle\psi|\sigma|\psi\rangle\geq2^{-\Omega}$. Choose a product eigenbasis for $\sigma$ and write $\sigma=\sum_k q_k|\Phi_k\rangle\langle\Phi_k|$, where $q_k\geq0$ and $\sum_k q_k=1$. Each $|\Phi_k\rangle$ has the form in Eq.~\eqref{EEI-saturation-state}. Thus
\begin{equation*}
2^{-\Omega}\leq\sum_k q_k|\langle\Phi_k|\psi\rangle|^2
\leq\max_k|\langle\Phi_k|\psi\rangle|^2.
\end{equation*}
Selecting a maximizing vector proves Eq.~\eqref{EEI-stability-overlap}. The trace distance between two pure states equals the square root of one minus their squared overlap. Together with $1-2^{-x}\leq x\ln2$, this proves Eq.~\eqref{EEI-stability-distance}.
\end{proof}

The estimate bounds distance to the exact EEI-saturating family, with no factor depending on the number or dimensions of the parties. In particular, $\Delta_{\mathrm{EEI}}\leq\varepsilon$ certifies a product of two pure end-pair states and pure interior states within trace distance $\sqrt{\varepsilon\ln2}$. This conclusion uses the stronger cycle deficit; approximate recoverability from a small chain deficit alone is a different statement. No optimality claim is made for the quantitative constant. The exact block decomposition thus turns the edge deficit into a quantitative estimate for its particular saturating family.

Table~\ref{tab:saturation-guide} summarizes the equality examples and structural characterizations.

\begin{table}[tbp]
\centering
\small
\setlength{\tabcolsep}{4pt}
\renewcommand{\arraystretch}{1.15}
\def\saturationcell#1#2{%
  \parbox[t]{#1}{\raggedright #2}%
}
\begin{tabular}{lll}
\hline\hline
\saturationcell{0.27\linewidth}{Inequality} &
\saturationcell{0.32\linewidth}{Equality example} &
\saturationcell{0.33\linewidth}{Qualification}\\
\hline

\saturationcell{0.27\linewidth}{EPI, MEI, SMEI, powered forms;\\ sharp $k$-marginal bound} &
\saturationcell{0.32\linewidth}{One pure entangled pair} &
\saturationcell{0.33\linewidth}{The distinguished cut separates the pair.}\\
\saturationcell{0.27\linewidth}{} &
\saturationcell{0.32\linewidth}{} &
\saturationcell{0.33\linewidth}{}\\[-5pt]

\saturationcell{0.27\linewidth}{Average upper bound} &
\saturationcell{0.32\linewidth}{$\mathrm{AME}(n,d)$} &
\saturationcell{0.33\linewidth}{Every order, whenever the state exists.}\\[5pt]

\saturationcell{0.27\linewidth}{Average marginal hierarchy} &
\saturationcell{0.32\linewidth}{Generalized GHZ state} &
\saturationcell{0.33\linewidth}{$\overline{\mathcal{E}}_r=H_2(p)$ at every proper order.}\\[5pt]

\saturationcell{0.27\linewidth}{$k$-MEI; TMEPI and GEPI} &
\saturationcell{0.32\linewidth}{Fully separable states} &
\saturationcell{0.33\linewidth}{Only such states attain equality at interior orders; see Example~\ref{ex:finite.range.bonds-sharp.and.weak.marginals-cyclic.bond.saturation}.}\\[15pt]

\saturationcell{0.27\linewidth}{CEI} &
\saturationcell{0.32\linewidth}{Independent adjacent pure bonds} &
\saturationcell{0.33\linewidth}{Complete characterization:
Proposition~\ref{prop:pure-chain-saturation}.}\\[15pt]

\saturationcell{0.27\linewidth}{HCEI} &
\saturationcell{0.32\linewidth}{Bonds of span at most $m$; open cluster states} &
\saturationcell{0.33\linewidth}{The cluster examples are genuinely multipartite entangled; see Example~\ref{ex:cluster-hyperchain}.}\\[15pt]

\saturationcell{0.27\linewidth}{EEI, HEEI, power forms} &
\saturationcell{0.32\linewidth}{One cyclically adjacent entangled pair} &
\saturationcell{0.33\linewidth}{The distinguished window separates the pair.}\\[15pt]

\saturationcell{0.27\linewidth}{Symmetric difference;\\ PHEPI and PGEPI} &
\saturationcell{0.32\linewidth}{Profiles $(0,h,h)$ and $(h,h,0)$} &
\saturationcell{0.33\linewidth}{Even and odd incidence:
Example~\ref{ex:parity-equalities}.}\\
\saturationcell{0.27\linewidth}{} &
\saturationcell{0.32\linewidth}{} &
\saturationcell{0.33\linewidth}{}\\[-5pt]

\saturationcell{0.27\linewidth}{WTPCEI, coefficient $\mu_{ij}$} &
\saturationcell{0.32\linewidth}{One pure entangled pair $(x,\ell)$, $x\in\{i,j\}$, $\ell\in R$} &
\saturationcell{0.33\linewidth}{$W(x,\ell)=\mu_{ij}$.}\\[15pt]

\saturationcell{0.27\linewidth}{WTPCEI, refined form} &
\saturationcell{0.32\linewidth}{At most three independent pure bonds; $|\mathrm{GHZ}_3\rangle$, $|\mathrm{GHZ}_5\rangle$} &
\saturationcell{0.33\linewidth}{Every realizable $(\mathcal{E}_{A_i},\mathcal{E}_{A_j},\mathcal{E}_{A_iA_j})$;\\ see Example~\ref{ex:weighted-two-party-ghz}.}\\[15pt]

\saturationcell{0.27\linewidth}{WOPCEI, coefficient $r_i$} &
\saturationcell{0.32\linewidth}{One pure entangled pair $(i,\ell)$, $\ell\neq i$} &
\saturationcell{0.33\linewidth}{$W(i,\ell)=r_i$; see Corollary~\ref{cor:weighted-one-party-cut}.}\\[5pt]

\saturationcell{0.27\linewidth}{HSEI; SEI for $n\leq4$} &
\saturationcell{0.32\linewidth}{A hub--partner entangled pair} &
\saturationcell{0.33\linewidth}{Suitable families attain the universal coefficients; for $n\geq5$, SEI equality forces a fully separable state.}\\[20pt]

\hline\hline
\end{tabular}
\caption{Equality cases for the inequalities in Sec.~\ref{sec:Ent.Ineq.}. All states are pure. Fully separable states saturate every listed homogeneous inequality.}
\label{tab:saturation-guide}
\end{table}

\subsection{Monogamy and Recoverability}
\label{subsec:monogamy}

The CEI is an entropy-sharing constraint. It does not directly compare the entanglement inside the mixed two-party marginals. For example, the two-party marginals of $|\mathrm{GHZ}_n\rangle$, $n\geq3$, are separable even though every $\mathcal{E}_{A_iA_j}$ equals one. Thus $\mathcal{E}_{A_iA_j}$, which quantifies entanglement across $A_iA_j\mid\overline{A_iA_j}$, must be distinguished from $\mathsf{E}_{A_i\mid A_j}(\varrho_{A_iA_j})$, which quantifies entanglement between $A_i$ and $A_j$ within the reduced two-party state. This distinction is also essential when comparing polygon relations with CKW-type monogamy \cite{CKW00,OV06}.

For a pure state, the CEI deficit can be rewritten exactly as
\begin{equation}\label{MI-chain-budget}
\Delta_{\mathrm{CEI}}=\sum_{j=1}^{n}\mathcal{E}_{A_j}
-\sum_{j=1}^{n-1}\mathcal{I}(A_j:A_{j+1}).
\end{equation}
The first sum is the total correlation of the globally pure state. Consequently, the deficit is the part left after subtracting the nearest-neighbor mutual information along the chosen path. Its nonnegativity does not hold for an arbitrary graph with cycles; the tree organization is essential for this particular correlation budget.

The entropy chain rule and SSA also give a tree extension of Eq.~\eqref{mixed-chain-entropy} \cite{PH11}. For any tree $T$ on $[n]$ and an arbitrary, possibly mixed state, root the tree and order its vertices so that each parent precedes its children. Replacing the full past in each conditional entropy by the parent gives
\begin{equation}
\Delta_T\coloneqq\sum_{\{i,j\}\in E(T)}\mathcal{S}(\varrho_{A_iA_j})-\sum_{j=1}^n\bigl(\deg_T(j)-1\bigr)\mathcal{S}(\varrho_{A_j})-\mathcal{S}(\varrho_{[n]})\geq0.
\end{equation}
For a pure state this is equivalently $\sum_{\{i,j\}\in E(T)}\mathcal{I}(A_i:A_j)\leq\sum_j\mathcal{E}_{A_j}$. The path recovers Eq.~\eqref{MI-chain-budget}; a star gives a different constraint involving its central marginal. The tree organization in this entropy bound has a different role from the incidence-parity condition used for the polygon inequalities.

When the one- and two-party entropies are known, the tree can itself be chosen to tighten the bound. Since $\Delta_T=\sum_j\mathcal{S}(\varrho_{A_j})-\mathcal{S}(\varrho_{[n]})-\sum_{\{i,j\}\in E(T)}\mathcal{I}(A_i:A_j)$, a maximum-weight spanning tree with weights $\mathcal{I}(A_i:A_j)$ minimizes $\Delta_T$. The maximum-spanning-tree choice has its classical antecedent in Chow--Liu approximation \cite{CL68}. Here the same optimization minimizes the quantum Markov entropy deficit; it does not by itself identify a closest quantum Markov state. This provides an explicit way of choosing a tree from the marginal data. It optimizes this particular entropy bound, without identifying a physical interaction network.

A precise connection to pairwise entanglement follows from squashed entanglement \cite{CW04}, defined for a bipartite state by
\begin{equation}\label{squashed-entanglement}
\mathsf{E}^{\mathrm{sq}}_{X\mid Y}(\varrho_{XY})
\coloneqq\frac{1}{2}\inf_{\omega_{XYR}:\,
\operatorname{Tr}_R\omega_{XYR}=\varrho_{XY}}
\mathcal{I}(X:Y\mid R)_\omega.
\end{equation}
The chain rule for conditional mutual information gives its monogamy relation $\mathsf{E}^{\mathrm{sq}}_{X\mid YZ}\geq\mathsf{E}^{\mathrm{sq}}_{X\mid Y}+\mathsf{E}^{\mathrm{sq}}_{X\mid Z}$: for every extension, decompose $\mathcal{I}(X:YZ\mid R)=\mathcal{I}(X:Y\mid R)+\mathcal{I}(X:Z\mid YR)$ and bound each term by its defining infimum. Taking a trivial extension also gives $2\,\mathsf{E}^{\mathrm{sq}}_{X\mid Y}\leq\mathcal{I}(X:Y)$. Together with the two exact expressions for the CEI deficit, this yields
\begin{align}\label{adjacent-squashed}
2\sum_{j=1}^{n-1}\mathsf{E}^{\mathrm{sq}}_{A_j\mid A_{j+1}} &\leq\sum_{j=1}^n\mathcal{E}_{A_j}-\Delta_{\mathrm{CEI}},\\ \label{skipping-squashed}
2\sum_{j=1}^{n-2}\mathsf{E}^{\mathrm{sq}}_{A_1\cdots A_j\mid A_{j+2}} &\leq\Delta_{\mathrm{CEI}}.
\end{align}
For the second inequality, the actual subsystem $A_{j+1}$ is an admissible extension of the reduced state on $(A_1\cdots A_j)A_{j+2}$. Each summand is therefore bounded by one half of the corresponding conditional mutual information. These are direct consequences of the definition and the deficit identity; they do not assert monogamy for the von Neumann entropy used as a mixed-state entanglement measure.

Equations~\eqref{adjacent-squashed} and \eqref{skipping-squashed} give complementary information. At fixed one-party entropies, the deficit reduces the upper bound on the sum of adjacent squashed entanglement, while bounding entanglement between the accumulated past and a site separated from it by its immediate predecessor. For a Bell-pair product on a matching of the path, the first bound is saturated and the second has zero on both sides.

Global purity also turns the last prefix conditional mutual information into ordinary mutual information: $\mathcal{I}(A_1\cdots A_{n-2}:A_n\mid A_{n-1})=\mathcal{I}(A_1\cdots A_{n-2}:A_n)$. Nonnegativity of the other terms and data processing therefore give
\begin{equation}
2\,\mathsf{E}^{\mathrm{sq}}_{A_1\mid A_n}
\leq\mathcal{I}(A_1:A_n)\leq\Delta_{\mathrm{CEI}}.
\end{equation}
Small deficit thus bounds even the total correlation between the path endpoints. A positive deficit need not represent quantum entanglement: the generalized GHZ example has classically correlated endpoints and zero endpoint squashed entanglement.

As an application of the established recovery theorem of Ref. \cite{FR15}, the conditional-mutual-information decomposition also gives an operational meaning to approximate saturation. To include ordinary chains and hyperchains in one statement, for an arbitrary state define
\begin{equation}
\Delta_m(\varrho)\coloneqq\sum_{j=1}^{n-m}\mathcal{S}(\varrho_{A_j\cdots A_{j+m}})-\sum_{j=1}^{n-m-1}\mathcal{S}(\varrho_{A_{j+1}\cdots A_{j+m}})-\mathcal{S}(\varrho_{[n]})=\sum_{t=m+2}^{n}c_t,
\end{equation}
where $c_t\coloneqq\mathcal{I}(A_1\cdots A_{t-m-1}:A_t\mid A_{t-m}\cdots A_{t-1})$. For a pure state $\varrho=|\psi\rangle\langle\psi|$, this definition gives $\Delta_m(\varrho)=\Delta_m$; in particular, $\Delta_1(\varrho)=\Delta_{\mathrm{CEI}}$.

\begin{proposition}[Sequential recovery from the hyperchain deficit]\label{prop:hyperchain-recovery}
For $1\leq m\leq n-2$, there are quantum channels
\begin{equation*}
\mathcal{R}_t:A_{t-m}\cdots A_{t-1}\longrightarrow A_{t-m}\cdots A_t, \qquad m+2\leq t\leq n,
\end{equation*}
whose sequential application to $\varrho_{A_1\cdots A_{m+1}}$ produces a state $\widetilde\varrho$ satisfying
\begin{equation}
\frac{1}{2}\|\varrho-\widetilde\varrho\|_1
\leq\min\Bigl\{1,\sum_{t=m+2}^{n}\sqrt{1-2^{-c_t}}\Bigr\}
\leq\min\Bigl\{1,\sqrt{(n-m-1)\ln2\;\Delta_m(\varrho)}\Bigr\}.
\end{equation}
Identity channels on the earlier sites are implicit.
\end{proposition}
\begin{proof}
Write $P_t=A_1\cdots A_t$ and $B_t=A_{t-m}\cdots A_{t-1}$. The recovery theorem of Ref. \cite{FR15} provides a channel $\mathcal{R}_t$ such that, with the root-fidelity convention $F(\rho,\sigma)=\|\sqrt\rho\sqrt\sigma\|_1$,
\begin{equation*}
F\bigl(\varrho_{P_t},
(\mathrm{id}_{P_{t-m-1}}\otimes\mathcal{R}_t)(\varrho_{P_{t-1}})\bigr)
\geq2^{-c_t/2}.
\end{equation*}
The Fuchs--van de Graaf inequality \cite{FG99} converts this fidelity bound into trace distance at most $\sqrt{1-2^{-c_t}}$. Start with $\widetilde\varrho_{P_{m+1}}=\varrho_{P_{m+1}}$ and define $\widetilde\varrho_{P_t}=(\mathrm{id}_{P_{t-m-1}}\otimes\mathcal{R}_t)(\widetilde\varrho_{P_{t-1}})$. By contractivity of trace distance, the error at step $t$ is at most its previous value plus $\sqrt{1-2^{-c_t}}$. Induction proves the first bound. Finally, $1-2^{-x}\leq x\ln2$ and Cauchy--Schwarz give the second.
\end{proof}

For the CEI, this gives an explicit reconstruction error of at most $\sqrt{(n-2)\ln2\;\Delta_{\mathrm{CEI}}}$. More generally, the hyperchain controls recovery with an $m$-site conditioning subsystem. The estimate is useful when $(n-m-1)\Delta_m\ll1$; the existence of a finite bond dimension alone supplies no such conclusion. The channels in the proposition are chosen for the given state, and their outputs need not preserve all previously specified marginals exactly.

At zero deficit every recovery step is exact. At small positive deficit the conclusion is closeness to a sequentially recovered state, not a distance bound to the set of exact quantum Markov states. These notions require care in quantum systems \cite{ILW08}. The inequality therefore has a precise reconstruction interpretation while retaining the distinction between exact conditional independence and approximate recoverability.

\section{Conclusion}\label{conclusion}

We have extended one-party entanglement polygon constraints to larger subsystem cuts and to families of cuts organized by graphs and hypergraphs. The marginal inequalities lead to an average hierarchy and a sharp $k$-marginal bound, derived directly by averaging strong subadditivity over both sides of a bipartition. Paths and cycles give chain and edge inequalities; overlapping hyperedges extend the chain construction, while uniform incidence parity gives polygon inequalities for general hypergraphs. For partial hyperstars, we determine the optimal coefficient that depends only on the number of parties, the marginal order, and the number of selected hyperedges. A particular prescribed hyperstar may admit a stronger coefficient. For a distinguished two-party cut and any weighted family of cuts, we determined the optimal polygon coefficient and, at fixed one- and two-party entropies of the distinguished pair, the exact minimum of the weighted sum. This criterion contains the cycle and complete-graph inequalities and shows that uniform incidence parity is not necessary for a polygon inequality. Adjoining a pure product party extends it to a distinguished one-party cut, and merging parties extends it to cuts not split by the family. Together with complementarity, the optimal weighted coefficient is therefore determined for every distinguished cut when $n\leq5$.

The exact chain and hyperchain deficits characterize equality through vanishing conditional mutual information terms. Pure chain equality states admit a representation by independent adjacent pure bonds embedded by local isometries; for qubits, the nontrivial bonds form a matching of the path. Open cluster states saturate the hyperchain inequality for overlaps of at least two sites despite having a positive chain deficit. For the cycle inequality, near equality gives a trace-distance bound to the exact equality family that is independent of the number of parties and their local dimensions. Applications provide necessary bond-dimension bounds for exact matrix product state representations, constraints on correlations and squashed entanglement, and sequential-recovery error estimates.

Polygon relations have been studied for other entanglement measures \cite{YYL22,Shi23,LGC24}. Which of the marginal, chain, and graph- or hypergraph-based inequalities established here remain valid for such measures? Answering this question requires identifying the properties used by each proof; validity of a one-party polygon inequality alone does not establish the higher-order bounds. Other questions include classifying equality for general parity hypergraphs and determining which of their polygon inequalities are sharp or redundant, and whether the optimal weighted coefficient for a distinguished cut $I_k$ with $3\leq k\leq n-3$ equals $m_{I_k}=\min_{u\in I_k,\,v\in\bar{I}_k}W(u,v)$. A single entangled pair shows that this value is always an upper bound; Theorem~\ref{theorem:weighted-two-party-cut} and Corollary~\ref{cor:weighted-one-party-cut} show that it is attained for $k\leq2$, and Proposition~\ref{prop:unsplit-weighted-cut} for families that do not split the cut. The first open case is $n=6$, $k=3$. For tensor networks, it would be useful to determine whether selected graph and hypergraph cuts give bond-dimension constraints beyond one-dimensional chains, and whether the present bounds can be made stable for approximate representations.

\begin{acknowledgments}
The author warmly thanks Karol \.Zyczkowski for several stimulating discussions, his encouragement, and for bringing Ref.~\cite{HSS03} to the author's attention. The author acknowledges financial support from the European Union under ``ERC Advanced Grant TAtypic, Project No.~101142236''. The views and opinions expressed are, however, those of the author only and do not necessarily reflect those of the European Union or the European Research Council Executive Agency. Neither the European Union nor the granting authority can be held responsible for them.

The author developed the research, carried out the analysis, and prepared the manuscript. Claude Opus 5.5 (by Anthropic) was used for assistance with drafting and revising text, checking mathematical arguments, bibliographic research, and preparing Figs.~\ref{fig:path-cycle},~\ref{fig:hypergraph-families},~\ref{fig:weighted-cut}. In particular, it brought Refs.~\cite{Pip03,FHC22,Filmus13} to the author's attention after the author's independent derivation of Lemma~\ref{lem:average-marginal-hierarchy} and Theorem~\ref{Theorem:SHMEI}. The author reviewed and verified the arguments, literature comparisons, figures, and final text, and takes full responsibility for the manuscript.

\end{acknowledgments}


\end{document}